\documentclass[a4paper,11pt]{article}
\usepackage{jheppub} 
\usepackage{lineno}
\usepackage{physics}
\usepackage{amsmath}
\usepackage{amssymb}
\usepackage{mathtools}
\usepackage{amsfonts}
\usepackage{bbold}

\title{\boldmath Natural disorder distributions from measurement}

\author[1,a,b]{Šárka Blahnik\note{Corresponding author.}}
\author[a,b]{Sarah Shandera}
\affiliation[a]{Institute for Gravitation and the Cosmos, The Pennsylvania State University, University Park, PA 16802, USA}
\affiliation[b]{Department of Physics, The Pennsylvania State University, University Park, PA, 16802, USA}

\emailAdd{seb6087@psu.edu}

\abstract{We consider scenarios where the dynamics of a quantum system are partially determined by prior local measurements of some interacting environmental degrees of freedom. The resulting effective system dynamics are described by a disordered Hamiltonian, with spacetime-varying parameter values drawn from distributions that are generically neither flat nor Gaussian. This class of scenarios is a natural extension of those where a fully non-dynamical environmental degree of freedom determines a universal coupling constant for the system. Using a family of quasi-exactly solvable anharmonic oscillators, we consider environmental ground states of nonlinearly coupled degrees of freedom, unrestricted by a weak coupling expansion, which include strongly quantum non-Gaussian states. We derive the properties of distributions for both quadrature and photon number measurements. Measurement-induced disorder of this kind is likely realizable in laboratory quantum systems and, given a notion of naturally occurring measurement, suggests a new class of scenarios for the dynamics of quantum systems in particle physics and cosmology.}

\begin{document}
\maketitle
\flushbottom

\section{Introduction}
The particle physics we understand in detail, the Standard Model, is known to encompass only a fraction of the degrees of freedom that are dynamically relevant in nature. While the beyond-Standard-Model particles are not (yet) directly experimentally accessible, they may impact the dynamics of the matter we see in a variety of ways. For example, the apparent constants in our model of the universe may be set by dynamical degrees of freedom that are pinned to some stable point for the times and energies accessible to us. Such dynamics has been invoked to determine the coupling constants in the Standard Model Lagrangian \cite{Appelquist:1987nr,Kerner:1968etm,Goldberger:1999} and the values of cosmological parameters \cite{Bousso:2000}. In many such scenarios, the extra degrees of freedom are stabilized in a simple state (usually the ground state, a pure state) and the resulting parameters in the effective theory take space- and time-independent values. While simple, calculable, and possibly the effective theory at work in the universe, this kind of construction is not the only possibility. Indeed, there is considerable interest in understanding the rich phenomenology when parameters vary with geometry and space or time \cite{Davoudiasl:2022kdq, Brzeminski:2023wza}, that is, in understanding effective field theories and cosmologies with disorder \cite{Tye:2007,Podolsky:2008du,Rothstein:2012hk,Amin:2015ftc, Green:2014xqa,Craig:2017ppp,Tropper:2020yew,Heckman:2022,Chaykov:2023}. 

When the effective description of a system contains disorder, interaction strengths and masses (or single-site energies in spin chains) take spatially or temporally varying values, drawn from some distribution \cite{Vojta:2019}. Most studies of disorder, beginning with the classic literature \cite{Anderson:1958}, use simple and sometimes ad hoc distributions to source the statistical variation. The two typical choices are flat and Gaussian, which can be well-motivated in some contexts \cite{Mckenzie:1996, Liu:2006, Aharony:2015aea, Makarov:2018}. For example, a flat distribution may be natural in scenarios where an ensemble of equal-energy (equally probable) geometric constructions determine the range of coupling constants considered. Gaussian distributions may be motivated when a nearly free quantum field sources the disorder. In addition, focus on simple distributions is reasonable as disordered systems are difficult to solve, and known examples already reveal qualitatively new phenomena \cite{Vojta:2019}. However, the simple cases may not be the most natural.

The notion of disorder as generated by a lack of complete information or control of an experimental set-up is well articulated in the literature (e.g., \cite{Kropf:2016}). In the lab, the experimentalist usually tries to suppress any effects of the unmeasured or uncontrolled environment, but a similar suppression need not happen for disorder phenomena in nature. In this paper, we seek to more fully characterize disorder that originates from the quantum state of the environment, through projective measurements on part of the environment. We conjecture that this is a plausible natural phenomenon, where a local process essentially equivalent to measurement determines aspects of the subsequent dynamics. 

If a system of interest is embedded in a very complex environment, where the measured fields are entangled with many other environmental fields, thermal statistics for disorder may be most typical. However, a typical environment often contains fields with a range of masses and charges, arranged in several semi-isolated sectors. In that case, only a subset of the degrees of freedom may be accessible in a typical measurement, and so only a subset will determine the disorder. For many reasonable environments, the result of a partial local measurement process is then a family of inhomogeneous or disordered Hamiltonians whose variation is governed by a non-flat distribution. Such an overall structure may also be expected when subsystems display out-of-equilibrium, or non-thermalized, behavior. Gaussian disorder arises in the special case of only Gaussian states and Gaussian measurements. 

An open-systems perspective on the origin of disorder is so far rather unexplored. A toy model for the effects of moduli coupling to light hidden-sector fields was recently considered in \cite{Balasubramanian2021}, but restricted to linearly coupled harmonic oscillators. Here we make a first foray into the broader class of systems that are not quadratic. To explore qualitative properties of nonlinearly coupled systems without the limitation of perturbation theory, we use a class of strongly anharmonic oscillators whose ground-state wave functions (and some excited states) are known exactly. Although we carry out detailed computations only for non-relativistic quantum systems rather than field theories, the notion should carry over to mode decompositions of a field in a rather straightforward way (eg, see \cite{Balasubramanian2021} for a discussion of this extension in the quadratic case). And, while the nonlinear systems considered here are specialized to be exactly solvable, the remarkable advances in engineering Hamiltonians for spin-chains \cite{Monroe:2021} and the broad interest in understanding thermodynamic phases of disordered systems provide a potential laboratory context for this general idea without such a restriction. We compute statistics for a simple, single quadrature measurement (a Gaussian measurement) and for the number operator, a non-Gaussian measurement relevant for photon number-detection schemes. 

In the next section we give a brief overview of the utility and description of non-Gaussian quantum states. We then characterize the quantum ground states of a family of quasi-exactly solvable anharmonic oscillators, which will be our example system to understand qualitative properties of disorder distributions beyond the Gaussian and thermal cases. In Section \ref{sec:CoupledOsc}, we consider systems of coupled oscillators and the statistics that would be obtained by measurements of (the mixed state of) just one oscillator. By using the quasi-exactly solvable family and assuming the existence of a decoupling frame for the oscillators, we are able to again compute disorder distributions from exact non-Gaussian states. We conclude with implications in Section \ref{sec:conclude}.

\section{Distributions from exact non-Gaussian pure states}
In this section, we define a family of anharmonic oscillator scenarios whose ground state wave functions are non-Gaussian. Because the anharmonic oscillators are quasi-exactly solvable, their exact ground states are known and so the distribution of measurement outcomes for any operator can also be calculated exactly. The class is particularly interesting because it contains cases that cannot be accurately captured in perturbation theory. 

\subsection{Characterizing non-Gaussian distributions}
\label{sec:characterize}
Classically, a non-Gaussian distribution for a variable $z$ is characterized by a complete set of moments, $\mu_n=\langle z^n\rangle$ for $n=0,1,2,3,...\,$. Since Gaussian distributions have non-zero even moments that are determined by the variance, $\sigma^2\equiv\langle z^2\rangle$, it is common to characterize non-Gaussian distributions in terms of cumulants, which are zero for $n>2$ for a Gaussian distribution. Non-Gaussian distributions can be written as expansions in the cumulants (eg, the Edgeworth series, an asymptotic expansion), but any truncation may not result in a proper probability distribution. Perturbative techniques may accurately generate the dominant non-Gaussian moments, usually appropriate for weak coupling in a Hamiltonian, but will fail at moderate and strong coupling. 

The distributions of quadrature statistics (e.g., position) from the anharmonic oscillators we will consider here are always symmetric, so only even moments are non-zero. We therefore characterize those distributions by the series of excess even moments
   \begin{equation}
       \nu_n \equiv  \frac{\mu_n}{\sigma^n} - (n-1)!!\;,\;\;\; n\;{\rm even}
   \end{equation}
where each moment is normalized by the appropriate power of the standard deviation, and the even (normalized) Gaussian moment of the same order is subtracted.

Quantum states are Gaussian if their phase space quasi-probability distribution, either the Wigner distribution or the Husimi Q-function, is Gaussian. Equivalently, measurement statistics of field quadratures are Gaussian for Gaussian states. Quantum states can be classically non-Gaussian, for example due to classical noise or probability mixtures. Quantum non-Gaussianity \cite{Walschaers:2021}, however, is more powerful. For example, it is the resource required to achieve a quantum advantage in continuous variable quantum computing \cite{Lloyd:1999,Albarelli:2018}. Here we will use the recently introduced stellar representation and stellar hierarchy \cite{Chabaud:2020} to characterize the degree of quantum non-Gaussianity of a state. These measures depend on the zeros of the Husimi Q-function \cite{Husimi1940SomeFP}, defined for a quantum state with density matrix $\rho$ by the trace of $\rho$ projected on the harmonic oscillator coherent state $|\alpha\rangle$ with complex amplitude $\alpha = \alpha_1 + i\alpha_2$,
\begin{equation}
\label{Qintro}
   Q(\alpha) =\frac{1}{\pi}\bra{\alpha} \rho |\alpha\rangle \,.
\end{equation}
A quantum non-Gaussian state is orthogonal to one or more coherent states. The stellar rank characterizes this in a concrete way, and is given by one half the number of zeros of the Q-function, counted with multiplicity \cite{Chabaud:2020}. Pure Gaussian states, and all non-Gaussian mixtures of Gaussian states, have no zeros and zero stellar rank. Fock states $|n_F\rangle$ have stellar rank $n$ (ie, $\alpha=0$ with multiplicity), and the cat and Gottesman-Kitaev-Preskill states are examples of infinite stellar rank \cite{Chabaud:2020}. While for pure states both negativity of the Wigner function and zeros of the Q-function diagnose quantum non-Gaussianity \cite{Lutkenhaus:1995}, for mixed states there exist positive Wigner function states that cannot be expressed as a convex combination of Gaussian states \cite{Filip:2011}. Since the set of states at each finite stellar rank are robustly separated, one can bound the degree of non-Gaussianity of a mixed state by computing the trace distance between it and a pure state of known stellar rank \cite{Chabaud:2021}. These considerations make the stellar rank an appropriate measure for our study, which includes mixed non-Gaussian states.

\subsection{Properties of quasi-exactly solvable sextic oscillators}
\label{sec:AnharmOsc}
In order to generate non-Gaussian pure and mixed states unrestricted by perturbation theory, we work with one of the established classes of anharmonic oscillators for which some exact results are known. We consider a family of sextic potentials \cite{Turbiner1988, Turbiner:2016} given by
\begin{equation}
\label{eq:potential}
    \tilde{V}_{\rm anharm}(n, k, \tilde{y}, a, b) = a^2\tilde{y}^6 + 2ab\tilde{y}^4 + [b^2-a(4n+2k+3)]\tilde{y}^2 - b(1 + 2k)\,.
\end{equation}
The cumbersome notation will be simplified shortly, via a reduction in the number of parameters and a useful rescaling. In Eq.(\ref{eq:potential}), $a \geq 0$ is a non-negative real number, $b$ is any real number, $k = 0$ or $1$, and $n$ is a non-negative integer. If $k=0$ the first $n$ eigenstates of even parity are known algebraically, namely the ground state, second excited state, and so on up through the $2(n-1)^{th}$ even excited state. If $k=1$, algebraic expressions exist for the first $n$ eigenstates of odd parity. The remaining wave functions cannot be expressed as analytic functions\footnote{However, recent work has significantly enlarged the space of exactly, and completely, solvable anharmonic oscillators to a class of sextic potentials generated by the Heun equation and related to those we use here \cite{Levai:2019, Truong:2021}.}, so these oscillators are called ``quasi-exactly solvable" (QES) \cite{Turbiner1988,Ushveridze:1994,Turbiner:2016}. Since a measurement-induced disorder scenario requires the environmental fields to be dynamically stable, we consider only ground states and so restrict to potentials with $k=0$. For simplicity, we will also set $n=0$. Then, we may work with the potential
\begin{equation}
\label{eq:potential2}
    \tilde{V}_{\rm anharm}(\tilde{y}, a, b) = a^2\tilde{y}^6 + 2ab\tilde{y}^4 + (b^2-3a)\tilde{y}^2 - b\,.
\end{equation}

To more clearly analyze the behavior of this family oscillators, it is useful to define a rescaling, $y=a^{1/4}\tilde{y}$, $c=\frac{b}{a^{1/2}}$ and consider the one-parameter family of functions
\begin{equation}
\label{eq:rescaledV}
    a^{-1/2}V(y,c)=y^6+2cy^4+(c^2-3)y^2-c\,.
\end{equation}
The Hamiltonian with canonical kinetic term (now in the conjugate momentum to $y$) and this potential may also be expressed in terms of raising and lowering operators. The expression is lengthy, but two notable properties are that it does not commute with $\hat{N}=\hat{a}^{\dagger}\hat{a}$, and it contains all terms up to cubic order in $\hat{N}$ and $\hat{K}_1\equiv\frac{1}{2}(\hat{a}^{\dagger}\hat{a}^{\dagger}+\hat{a}\hat{a})$, with coefficients constrained by the form of Eq.(\ref{eq:rescaledV}).

\begin{figure}[htbp]
    \begin{centering}
    \includegraphics[width=.95\textwidth]
    {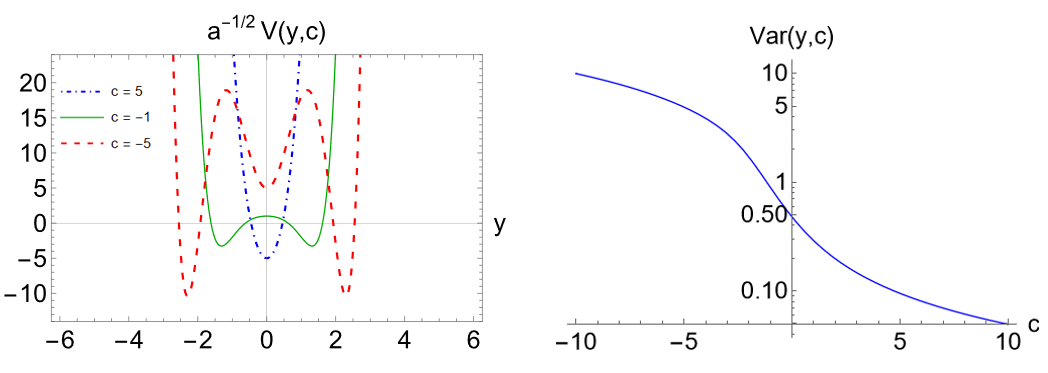}
    \end{centering}
      \caption{\textbf{Left:} The three classes of potential. The potential is single-welled for $c\geq\sqrt{3}$ (e.g. c = 5 in dot-dashed blue), double-welled for $-\sqrt{3}\leq c< \sqrt{3}$ (e.g. c = -1 in solid green), and a triple-welled when $c<-\sqrt{3}$ (e.g. c = -5 in dashed red). \textbf{Right:} Log-scale plot of the ground state variance of $y$ as a function of $c$. 
      \label{fig:potVar}}  
\end{figure}

One-dimensional anharmonic potentials of the form of Eq.(\ref{eq:rescaledV}) (or Eq.(\ref{eq:potential2})) can be classified into three types, according to whether they have one, three, or five extrema. This results in a sextic triple-well potential if $c<-\sqrt{3}$, a sextic double-well potential if $-\sqrt{3}\leq c < \sqrt{3}$, and single sextic well if $c \geq \sqrt{3}$. The left panel of Figure \ref{fig:potVar} shows an example of each possible shape. For all values of $c$, the ground state wave function for this class of oscillators is 
\begin{equation}
\label{eq:anhamronicWF}
\psi_0(y,c)=A(c)\exp\left[-\frac{1}{4}y^4-\frac{1}{2}cy^2\right]\,,
\end{equation}
which are clearly non-Gaussian. The normalization constant $A$ is a piecewise-defined function of $c$, with form dependent on whether $c$ is positive or negative (given by Equation \ref{eq:Atnorm} in Appendix A). The sign of $c$ is determined exclusively by the sign of $b$, as allowing $a$ to be negative would destroy the quasi-exact solvability. 

The moments of the ground state wave function are calculated by $\mu_n=\int_{-\infty}^{\infty} dy \, y^n \, \psi^*(y)\psi(y)$. Analytic expressions are given in Appendix \ref{app:analytic}. Note that the translation back to the original variables is straightforward for the moments, since $\expval{y^{2n}} = a^{\frac{n}{2}}\expval{\tilde{y}^{2n}}$, while the normalized excess moments are the same for the original and rescaled variables, $\nu_{2n}(y, c) = \nu_{2n}(\tilde{y}, a, b)$. As shown in the right panel of Figure \ref{fig:potVar}, the variance tends to zero for large positive $c$, as the single U-shaped potential narrows. At large negative $c$, the separation between left and right minima grows (even as each individual well becomes very locally narrow) and so the variance increases. 

\begin{figure}[htbp]
   \includegraphics[width=.4\textwidth]{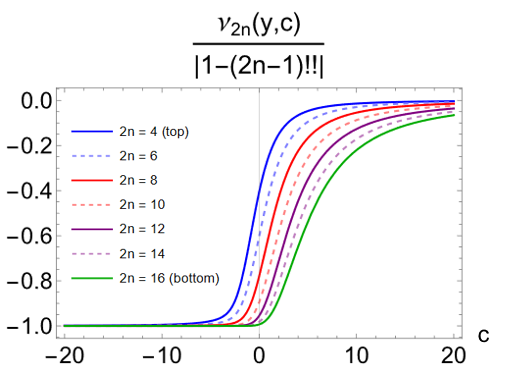}
   \includegraphics[width=.4\textwidth]{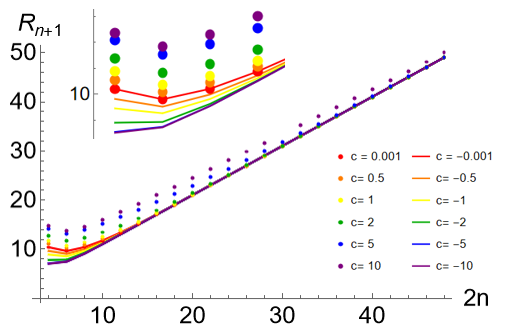}
    \caption{\textbf{Left:} Excess $2n^{th}$ even moments 4 through 16 as a function of $c$, with all curves rescaled via division by the absolute value of their respective lower bounds. \textbf{Right:} Successive even moment ratios $R_{n+1}(y,c)$, Eq.~(\ref{eq:momentRatio}), as a function of $2n$. Different curves correspond to different values of $c$, with smooth lines for negative $c$ and dotted lines for positive $c$. In both cases, the ratios tend towards $2n+1$, independent of the value of $c$. Lower moment ratios exhibit greater dependence on $c$ values, especially when $c>0$ (as shown in the inset enlargement of moments $2n=4,6,8,10$).  \label{fig:scurvesSuccessiveRatios}} 
\end{figure}

The higher order moments also have simple limiting behavior. Using the expressions in Appendix \ref{app:analytic}, we find
\begin{align}
    \lim_{c\to + \infty} \nu_{2n}(y,c) &= \frac{2^n\Gamma[n+\frac{1}{2}]}{\sqrt{\pi}} - (2n-1)!! \nonumber\\
    &= 0\,.
\end{align}
That is, all excess even moments approach zero at large positive $c$. While individually the moments approach those of a Gaussian in this limit, higher order moments are always larger than the lower order moments. At large, negative $c$, on the other hand, the distribution becomes extremely non-Gaussian. The standardized even moments, $\frac{\mu_n}{\sigma^n}$, all approach 1 as $c\rightarrow -\infty$ so that
\begin{equation}
    \lim_{c\to - \infty} \nu_{2n}(y,c) = 1 - (2n-1)!!\,.
\end{equation}
The left panel of Figure \ref{fig:scurvesSuccessiveRatios} shows the behavior the excess moments as a function of $c$.

Some additional structure of the non-Gaussianity can be clarified by looking at the relative importance of contributions from higher moments. The right panel of Figure \ref{fig:scurvesSuccessiveRatios} shows the ratios of two successive even moments, 
\begin{equation}
\label{eq:momentRatio}
    R_{n+1}(y,c) \equiv \frac{\nu_{2(n+1)}(y,c)}{\nu_{2n}(y,c)}\,.
\end{equation}
Even for $c\gg0$, the higher order moments are larger than lower order (they approach zero more slowly). In fact, the ratio tends to $2n+1$ for sufficiently large $n$, regardless of the value of $c$. This result is consistent with the discussion of the special point $a=0$ ($c=\infty$ in the rescaled variables) in \cite{Turbiner:2016}. There, it is noted that at small $a$ (large $c$) the quartic term dominates over the sextic. As there are no real or Hermitian fourth order Hamiltonians that are quasi-exactly solvable, the set of solutions does not have a smooth limit to the truly Gaussian point by taking $a\rightarrow0$ ($c\rightarrow\infty$). The behavior for negative $c$ is qualitatively similar to the ratios for positive $c$, but the lines merge together most rapidly when $c$ is negative. 

\begin{figure}[hbtp!]
    \centering
    \includegraphics[width=6.4 in]{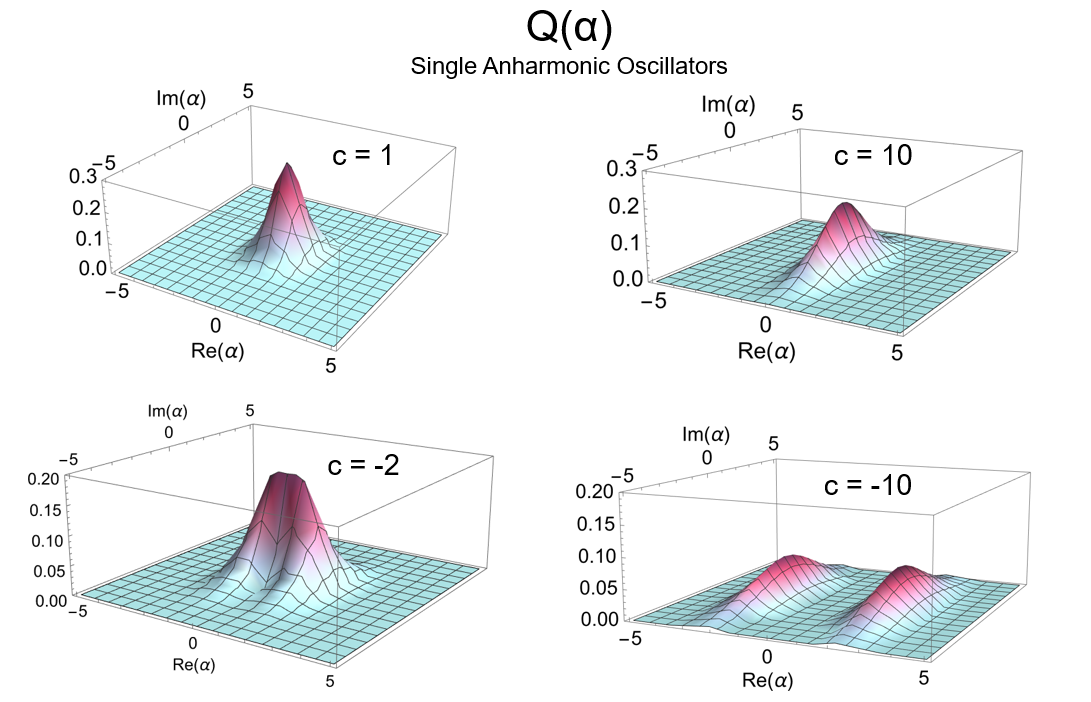}
    \caption{Husimi Q-functions for anharmonic oscillator states in the single-well regime of the potential ($c = 10$ in Eq.(\ref{eq:rescaledV})), the double-well regime ($c=1$), and the triple-well regime ($c=-2, -10$). For positive $c$ there is a single, central peak, which flattens and spreads out along the imaginary axis as $c\rightarrow+\infty$. As $c$ becomes increasingly negative, the central peak splits into two. These move apart from each other, with the imaginary axis as the central dividing line, flattening and spreading along the imaginary axis.
    \label{fig:Qfuncs1D}
    }
\end{figure}

Beyond the non-Gaussianity of the single-quadrature moments, we now consider the full quantum non-Gaussianity of these oscillators. Using 
\begin{equation}
   Q(\alpha) =\frac{1}{\pi}\bra{\alpha} \rho \ket{\alpha} 
    = \frac{1}{\pi}\int dy \, dy'
    \psi_{\alpha}^*(y) \rho(y, y') \psi_{\alpha}(y')
\end{equation}
and $\psi_{\alpha}(y)=\pi^{-1/4}e^{-(y-\sqrt{2}\alpha_1)^2/2}e^{i\sqrt{2}\alpha_2 y}$, with $\alpha=\alpha_1+i\alpha_2$, the Q-function for the ground state is given by
\begin{equation}
\label{eq:Q1D}
    Q(\alpha) = \frac{(A[c])^2}{\pi^{3/2}}e^{-2\alpha_1^2}|G_c(\alpha)|^2 \,,
\end{equation}
where
\begin{equation}
\label{eq:Gc}
G_c(\alpha) \equiv \int_{-\infty}^\infty dy \, \exp[-\frac{1}{4}y^4 - \frac{c+1}{2}y^2+\sqrt{2}\alpha_1 y]e^{i\sqrt{2}\alpha_2y}\,.
\end{equation}
The Q-functions for the ground state of the anharmonic potential with several values of $c$ are shown in Figure \ref{fig:Qfuncs1D}. 

\begin{figure}[htb]
   \centering
    \includegraphics[width=.9\textwidth]{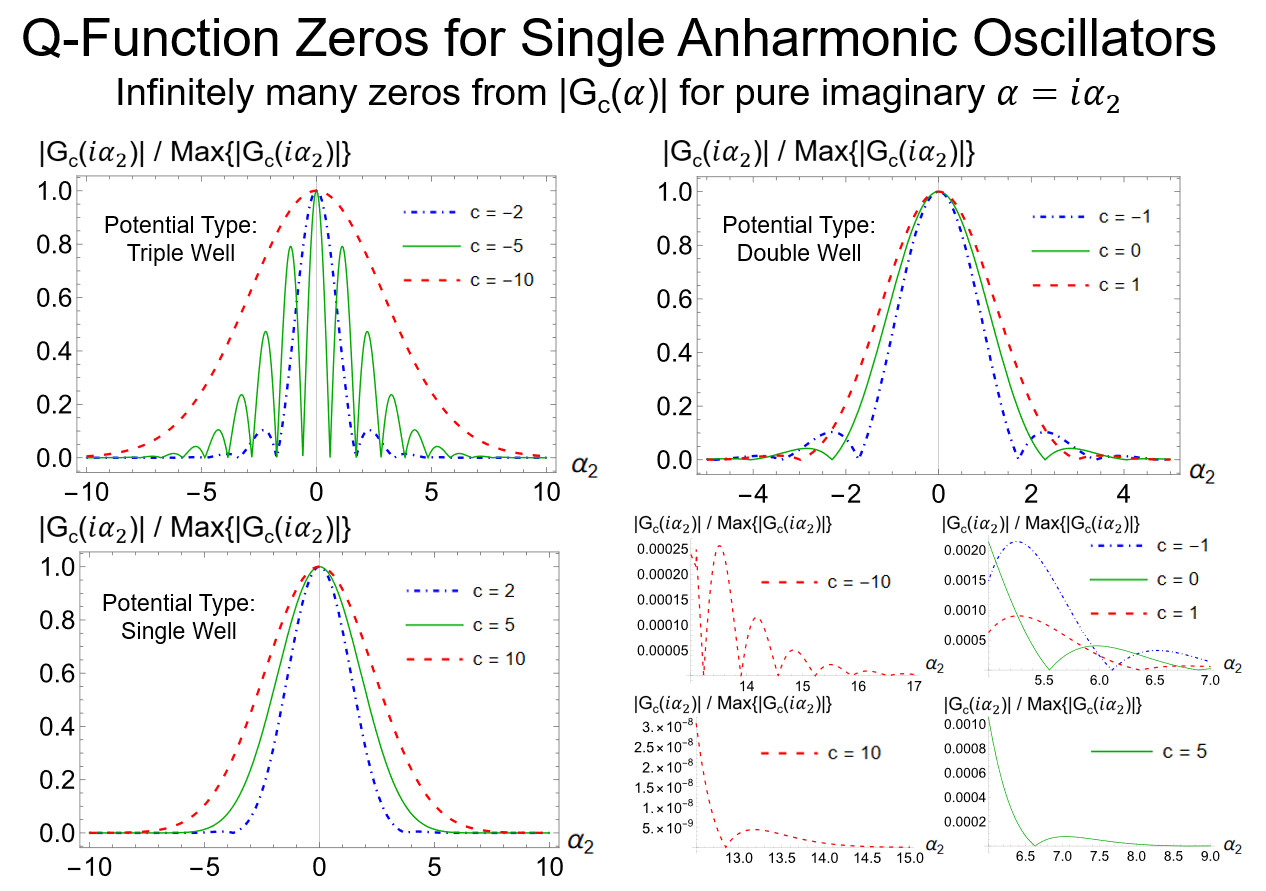}
    \caption{Plots of $|G_c(i\alpha_2)|$ (see Eq.(\ref{eq:Gc})) divided by its $c$-dependent maximum value, which always occurs at $\alpha = 0$ and ranges here from about $3.72 \times 10^8$ for $c=-10$ down to $0.751$ for $c = 10$. Examples for potentials in the triple well (top left), double well (top right) and single well (lower left) are all shown. The bottom right panel has a zoom in to show the continued small-amplitude oscillatory behavior out to large $\alpha_2$. In that region, the more strongly non-Gaussian states display a higher density of zeros.\label{fig:GcAllPotShapes}}
\end{figure}

Though the exponential damping term in Eq.(\ref{eq:Q1D}) ensures $Q(\alpha)$ will be very small whenever $\Re(\alpha)$ is large, the true zeros of this Q function occur only where $|G_c(\alpha)|^2 = 0$. If $\alpha_1\neq0$, neither the real or imaginary parts of $G_c$ will be separately zero, and so $\Re[G_c]^2+\Im[G_c]^2\neq0$. But, along $\alpha_1=0$, $\Im[G_c]=0$ by symmetry, and so a zero of the Q-function exists if $\Re[G_c]=0$. Since the cosine is an even function, zeros can only occur when the exponential damping suppresses the domains of positive integrand just enough more than those of negative integrand that they cancel. Although we are unable to provide an analytic expression for $G_c$, numerical evidence suggests that zeros do indeed occur, and there are likely to be infinitely many.  One may confirm numerically that $G_c$ indeed oscillates in sign at apparent zeros, and it seems likely the 1D anharmonic oscillator ground states are always of infinite stellar rank. This is true for both positive and negative $c$, although the distribution and density of the zeros along the $\alpha_2$ axis are dependent on $c$. For examples, see Figure \ref{fig:GcAllPotShapes}. 

While the exact functional dependence is not known, our numerical results indicate that the overall density of Q-function zeros increases as $c$ decreases. That is, states which are more non-Gaussian with respect to position operator moments also have a higher density of zeros. This trend suggests it may be interesting to consider if there is a refinement of the relative level of quantum non-Gaussianity even among states of infinite stellar rank. A related study comparing the degree of non-classicality with the degree of non-linearity in the potential, for polynomial potentials with small quartic and sextic terms, was carried out in \cite{Albarelli2016}.

\begin{figure}[htbp]
\begin{centering}
  \includegraphics[width=.4\textwidth]{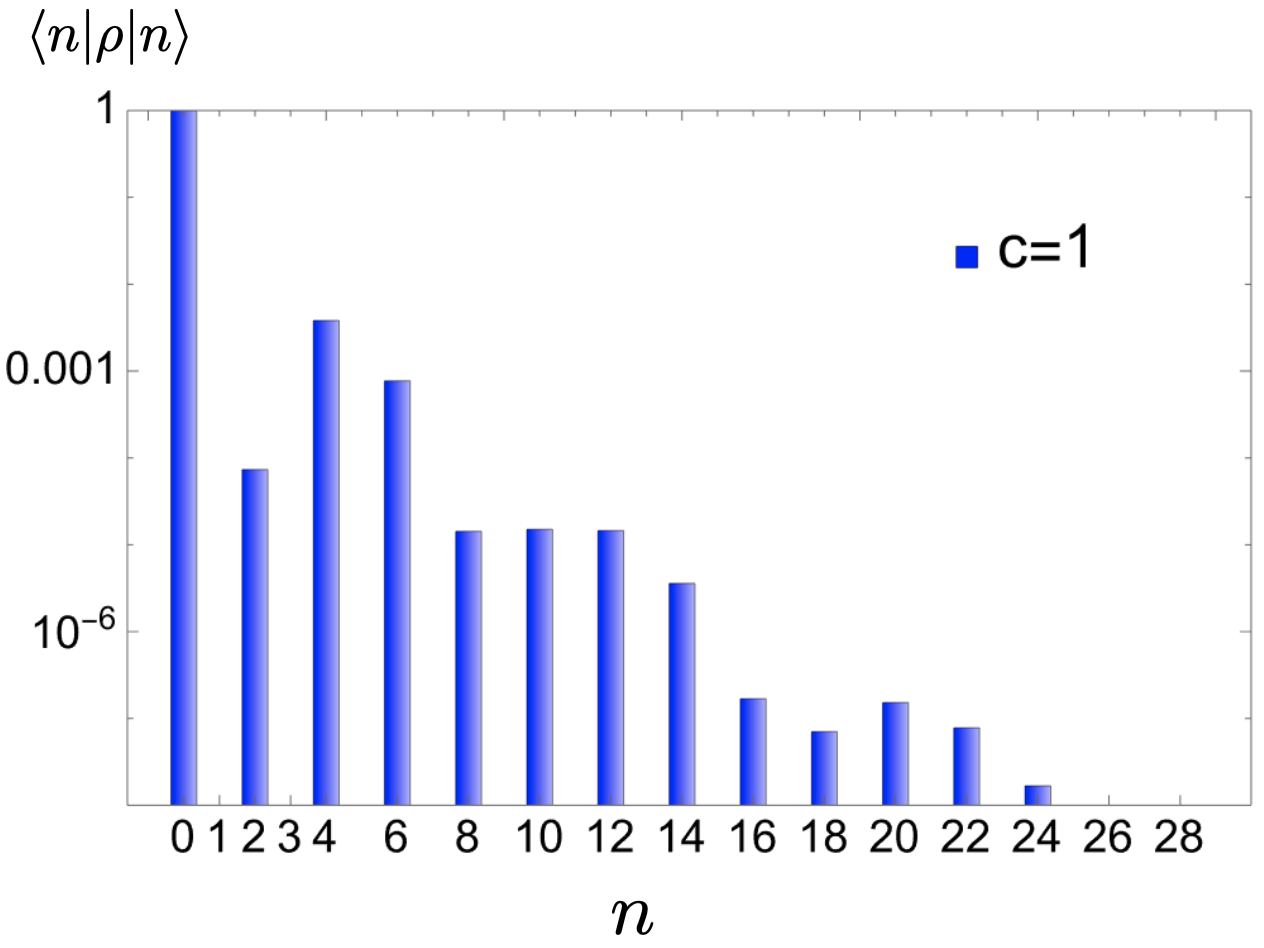}
  \includegraphics[width=.4\textwidth]{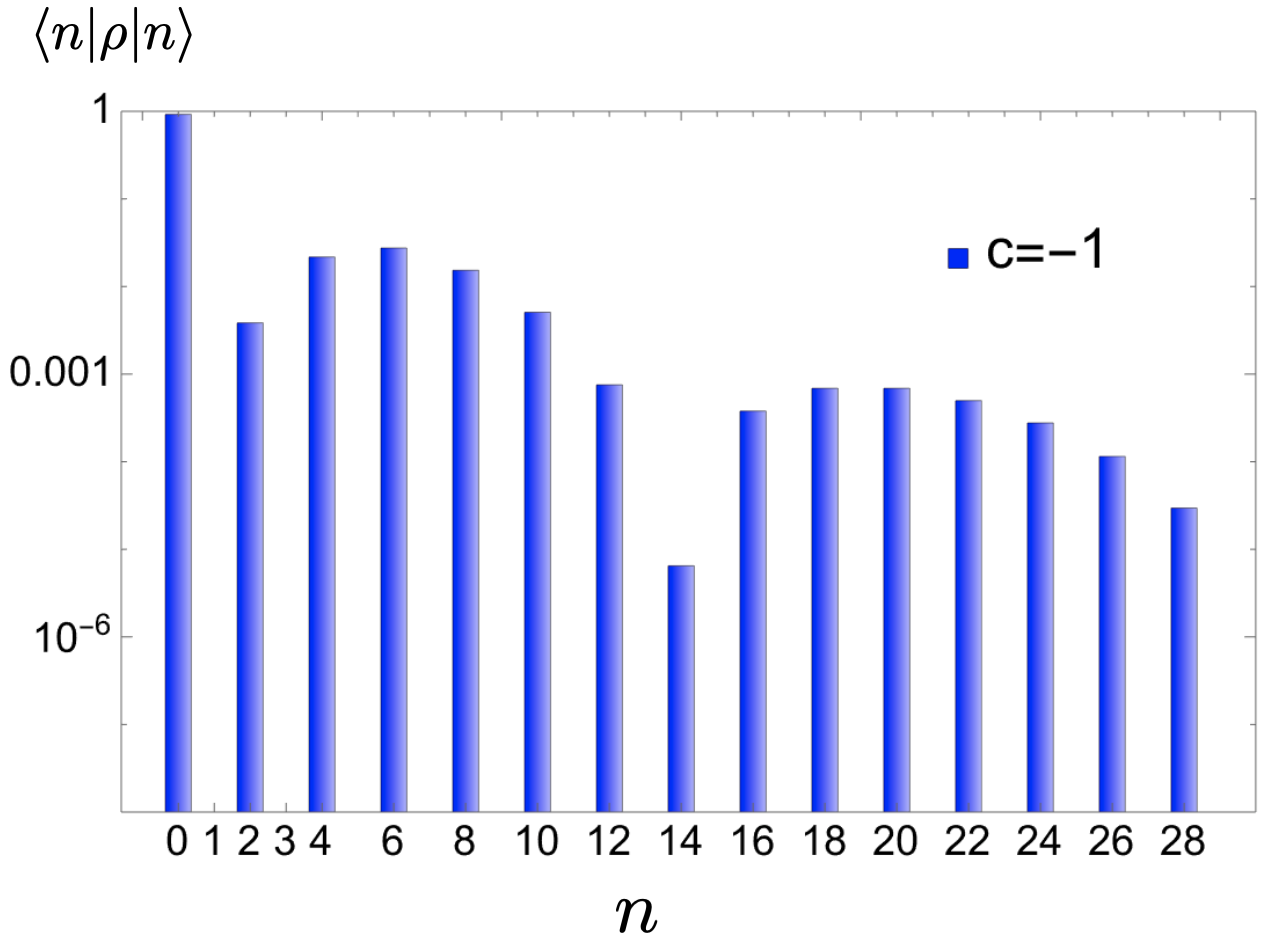}
    \caption{The number operator statistics for a single anharmonic oscillator with $c=1$ (left) and $c=-1$ (right). \label{fig:anharmonicNpops}}
\end{centering}
\end{figure}

Measurements of position are of course not the only thing of interest, or even the most natural in determining disorder. For single oscillators in the ground state, energy measurements do not generate disorder. However, since for these anharmonic systems energy eigenstates are not Fock states, the number operator, $\hat{N}=\hat{a}^{\dagger}\hat{a}$, provides another interesting operator to consider. The statistics of the number operator for modes of frequency $\omega$ may be computed for any density matrix via 
\begin{equation}
\label{eq:nstats}
   \langle n|\hat{\rho}|n\rangle=\left(\frac{\omega}{\pi}\right)^{1/2}\frac{1}{2^n n!} \int dx\,dx^{\prime}\,\rho(x,x^{\prime})e^{-\frac{\omega}{2} (x^2+x^{\prime 2})}H_n(x\sqrt{\omega})H_n(x^{\prime}\sqrt{\omega})\,,
\end{equation}
where $H_n(x)$ are the Hermite polynomials. We may define a useful frequency via the variance $\langle x_1^2\rangle\equiv \frac{1}{2\Omega_T}$ and carry out the integration above. For these symmetric anharmonic oscillators in the ground state, Eq.(\ref{eq:anhamronicWF}), only modes with $n$ even will have non-zero occupation. In general, states with more non-Gaussianity in the position-measurement distribution have more significant population at higher $n$. Figure \ref{fig:anharmonicNpops} shows two examples, $c=\pm1$. For triple-well potentials, $c<-\sqrt{3}$, there is lower occupation number in $n=22$ rather than $n=18$ dip seen for the double well cases. The occupation numbers in $n=0$, $n=2$, and $n=22$ all drop significantly as $c$ becomes more negative. As $c$ increases for $c>0$, the distribution becomes very strongly peaked about $n=0$, consistent with the (non-smooth) approach to a Gaussian ground state in this limit. 

\section{Distributions from mixed states}
\label{sec:CoupledOsc}
In this section, we characterize example distributions for disorder from partial measurements of an environment. We first construct the environmental state: the ground state of systems of two coupled anharmonic oscillators. Then we characterize the reduced state of a single oscillator with the other traced out. These examples will provide prototypical distributions for disorder from partial measurements of an environment. As a laboratory realization of the type of scenario we have in mind, consider engineering a disordered spin chain with the on-site magnetization set by the outcome of a photon count from a suite of identical neighboring cavities. If each cavity supports multiple, coupled modes, but only one of those is measured and its photons turn the dial generating the local magnetic field, then the distribution of magnetizations is given by the statistics of the number operator, $\hat{N}$, in the mixed state of a single mode. Although the example above is engineered, a typical ``environment" in beyond-Standard Model constructions contains more than one isolated degree of freedom. The observable degrees of freedom typically couple to only some (or even one) of those background fields, which mediate the interaction with the additional environmental fields. As described in \cite{Balasubramanian2021}, one might have in mind that the mediating environmental degree of freedom is a moduli field that couples to both the Standard Model and to some other hidden sector with light fields. Then, if a relevant dynamical process locally projects the mediating environmental degree of freedom onto a basis state, the effective Hamiltonian for the observable system will depend on the result of that process, (e.g., $\langle \hat{N}\rangle$), which will vary on the scale of the local projective process. The disorder distribution will be inherited from the effective measurement outcomes of a mixed state.  

\subsection{Exact results via a decoupling frame}
\label{sec:bilinear}

Typically, the study of coupled degrees of freedom must be carried out perturbatively. However, we can construct a class of (strongly) coupled anharmonic oscillators whose joint ground state is known exactly by assuming a decoupling frame. In other words, we invert the standard diagonalization process and begin with Hamiltonians of the form
\begin{align}
\label{eq:decoupling}
H =&H_1(y_1)+H_2(y_2)\,
\end{align}
where $H_1$ and $H_2$ are such that at least the ground states of each can be found exactly. (Depending on the degree of solvability of the $H_i$, additional states may also be known.) Then, the family of coupled oscillator systems whose ground state is also known exactly is obtained by a rotation of variables. In the simplest case of canonical kinetic terms in all frames, we need only define the coordinates $y_i$ to be given by a rotation of coordinates $x_i$:
\begin{equation}
\label{eq:rotdef}
	\begin{pmatrix}
		y_1  \\
		y_2 \\
	\end{pmatrix}
	=
	\begin{pmatrix}
		\cos\theta & -\sin\theta \\
		\sin\theta & \cos\theta \\
	\end{pmatrix}
	\begin{pmatrix}
		x_1  \\
		x_2 \\
	\end{pmatrix}\,.
\end{equation}
These unitary rotations define a continuous family of meronomic frames \cite{Hulse:2019}, parameterized by $\theta$, in which oscillators $\{p_{x_1},x_1\}$ and $\{p_{x_2},x_2\}$ are coupled. The choice of frame is completely irrelevant in the absence of something outside this system that depends on or determines it. In the present context of disorder from measurement, the measurement device or procedure picks out a particular $\theta$ and defines the entanglement of the reduced state of the relevant degree of freedom with its unmeasured environment. 

Below, we first review the familiar case of two linearly coupled harmonic oscillators and then construct specific examples of coupled anharmonic oscillators and examine the properties of the resulting single-oscillator reduced state. 

\subsection{Linearly coupled harmonic oscillators}
\label{sec:bilinear}
As a warm-up, we briefly review a few features of the simplest example: two linearly-coupled harmonic oscillators. This quadratic system was used in \cite{Balasubramanian2021} as an example of potential open-system effects on measurable degrees of freedom in particle physics or cosmology and is well-studied in many contexts, including quantum optics \cite{Weedbrook:2012,Jellal:2011tx,Adesso:2005,Adesso2007,Ferraro:2005} and early literature on the entropy of black holes \cite{Bombelli:1986,Srednicki:1993im}. 

The Hamiltonian for two harmonic oscillators in a decoupling frame is
\begin{equation}
    H=\frac{1}{2}p_{y_1}^2 +\frac{1}{2}\omega^{\prime 2}_1y_1^2 +p_{y_2}^2 + \frac{1}{2}\omega_2^{\prime 2}y_2^2\,.
\end{equation}
After changing frame by a rotation defined by angle $\theta$ in Eq.(\ref{eq:rotdef}), the Hamiltonian is
\begin{equation}
\label{H}
    H = \frac{1}{2}(p_{x_1}^2 + p_{x_2}^2) +
    \frac{1}{2}(\omega_1^2x_1^2+\omega_2^2x_2^2+\lambda x_1x_2)\, ,
\end{equation}
where $\omega_1^2=\omega_1^{\prime 2}\cos^2\theta+\omega_2^{\prime 2}\sin^2\theta$, $\omega_2^2=\omega_1^{\prime 2}\sin^2\theta+\omega_2^{\prime 2}\cos^2\theta$, $\lambda=2\cos\theta\sin\theta(\omega_2^{\prime 2}-\omega_1^{\prime 2})$, and $|\lambda|\leq 2\omega_1\omega_2$.

The ground state in the uncoupled frame is a product of the two single oscillator ground states,
\begin{equation}
    \psi_0(y_1,y_2)=\frac{(\omega_1^{\prime}\omega_2^{\prime})^{1/4}}{\pi^{1/2}}\exp\left[-\frac{1}{2}(\omega_1^{\prime}y_1^2+\omega_2^{\prime}y_2^2)\right]\,,
\end{equation}
and the ground state in the coupled frame is found by applying the same rotation used to generate the Hamiltonian:
\begin{equation}
    \psi_0(x_1,x_2)\propto\exp\left[-\frac{1}{4}\left(\frac{x_1^2}{\tau_1^2}+\frac{x_2^2}{\tau_2^2}+\frac{2x_1x_2}{g}\right)\right]\,.
\end{equation}
Here $(2\tau_1^2)^{-1}=\omega_1^{\prime}\cos^2\theta+\omega_2^{\prime}\sin^2\theta$, $(2\tau_2^2)^{-1}=\omega_2^{\prime}\cos^2\theta+\omega_1^{\prime}\sin^2\theta$, and $(2g)^{-1}=(\omega_2^{\prime}-\omega_1^{\prime})\sin\theta\cos\theta$. The state of just a single oscillator, $x_1$, is found by tracing out $x_2$. That is,
\begin{align}
    \label{reducedrhodef}
     \rho(x_1,x_1') =& \int dx_2\,\psi_0(x_1,x_2)\psi_0^{*}(x_1^{\prime}, x_2)\\\nonumber
 =&\sqrt{\frac{\gamma-\beta}{\pi}}\exp\left[-\frac{\gamma}{2}(x_1^2+x_1^{\prime 2})+\beta x_1x_1^{\prime}\right]
\end{align}
where $\gamma=\frac{1}{2\tau_1^2}-\frac{\tau_2^2}{4g^2}$, $\beta=\frac{\tau_2^2}{4g^2}$. 

The reduced state is still a Gaussian state, with mean $\langle x_1\rangle = 0$ and variance
\begin{align}
    \label{expvalfromrhored}
    \langle x_1^2\rangle 
    =&\,\tau_1^2\left(1 - \frac{\tau_1^2\tau_2^2}{g^2}\right)^{-1} =\frac{1}{4(\gamma^2-\beta^2)}\,.
\end{align}
The purity of the state is 
\begin{equation}
    \text{Tr}(\rho^2(x_1,x_1^{\prime})) = \sqrt{1-\frac{\tau_1^2\tau_2^2}{g^2}}
    = \sqrt{\frac{\tau_1^2}{\text{Var}(x_1)}}\,,
\end{equation}
which for Gaussian states is related to the covariance matrix, $\Sigma$, by $\Tr[\rho^2]=\frac{1}{2\sqrt{\det[\Sigma]}}$. From these expressions, it is clear that the variance of $x_1$ depends on the coupling and can be made very large when $g\approx\tau_1\tau_2$ \cite{Balasubramanian2021}. If one ignored the possibility of a more complex environment and attributed measurement results to a single field, the conclusions about the parameters of that field would be incorrect. In addition, a large variance and large coupling $\lambda$ lead to a higher degree of mixedness for the reduced state. The purity is plotted in the left panel of Figure \ref{fig:purityfuncofc2idpiov4}, where it can be contrasted with a simple example of the anharmonic case. 

Projective measurements of position are Gaussian operations, returning Gaussian statistics for any Gaussian state. One may also consider the simplest non-Gaussian measurement, photon counting. For pure Gaussian states, which are states that saturate the uncertainty relationship, the statistics of the number operator can be straightforwardly derived from the statistics of a single quadrature. Gaussian mixed states can always be related to a thermal state \cite{Williamson:1936, Weedbrook:2012}, and in this example there is a natural way to identify the thermal character. The ground-state variance defines a harmonic oscillator frequency via $\langle x_1^2\rangle\equiv \frac{1}{2\Omega_T}$, where 
\begin{equation}
    \Omega_T=2(\gamma^2-\beta^2)\,.
\end{equation}
With respect to this frequency, the state $\rho(x_1,x_1')$ is a thermal state \cite{Srednicki:1993im} at temperature 
\begin{equation}
    T=\frac{\sqrt{\Omega_T}/2}{\log\left(\frac{\gamma+\sqrt{\Omega_T}/2}{\beta}\right)}\,,
\end{equation}
and with mean occupation
\begin{equation}
    \langle N_T\rangle=\frac{\beta}{\gamma-\beta+\sqrt{\Omega_T}/2}\,. 
\end{equation}
A different choice of frequency corresponds to a different basis for the density operator, where it would not be diagonal. However, in any basis the density matrix can be related to a thermal state by applying a combination of squeezing and displacement operators \cite{Williamson:1936, Weedbrook:2012}. In all cases, the Gaussian nature of the system restricts the number operator statistics to have a particular shape \cite{Marian:1993}, with the distribution falling off at larger occupation number (inherited from the thermal part), but with oscillations whose frequency is determined by the squeezing parameter.

In summary, if measurements of one of a pair of bilinearly coupled oscillators, in the joint ground state, are used to determine subsequent parameters for some system, the most natural resulting disorder distributions are either Gaussian, exactly thermal, or related to the thermal distribution by squeezing and displacement. We will find a much broader class of distributions possible from anharmonic oscillators with non-Gaussian ground states.

\subsection{Coupled anharmonic oscillators}
\label{sec:AnharmCoupling}
An interesting class of non-Gaussian mixed states can be found by carrying out the same procedure as in the previous section, but for a pair of quasi-exactly solvable anharmonic oscillators. The general potential for two uncoupled sextic oscillators is
\begin{equation}
\label{eq:potential2d}
    V(\tilde{y}_1, \tilde{y}_2) = a_1^2\tilde{y}_1^6 + a_2^2\tilde{y}_2^6 + 2a_1b_1\tilde{y}_1^4 + 2a_2b_2\tilde{y}_2^4 +(b_1^2-3a_1)\tilde{y}_1^2 +(b_2^2-3a_2)\tilde{y}_2^2- (b_1+b_2)\,.
\end{equation}
By restricting to $a_1=a_2\equiv a$, we can define a common rescaling $c_i=\frac{b_i}{a^{1/2}}$, $y_i = a^\frac{1}{4}\tilde{y}_i$, and similarly $x_i = a^\frac{1}{4}\tilde{x}_i$ after rotating to any frame (using Eq.(\ref{eq:rotdef})) where the oscillators are coupled. The ground state wave function is then
\begin{multline}
\label{eq:psi0xtgen}
   \psi_0(x_1, x_2) = A_1A_2\exp{-\frac{c_1}{2}\left[(\cos\theta) x_1 - (\sin\theta) x_2\right]^2
    -\frac{c_2}{2}\left[(\sin\theta) x_1 + (\cos\theta)x_2\right]^2}\\
   \times \exp{-\frac{1}{4}\left[(\cos\theta) x_1 - (\sin\theta)x_2\right]^4 -\frac{1}{4}\left[(\sin\theta) x_1 + (\cos\theta) x_2\right]^4}\,.
\end{multline}
Appendix \ref{app:analyticcoupled} shows the class of sextic, coupled Hamiltonians that are consistent with the decoupling frame, and contains some additional results without the rescaling. Equation (\ref{eq:psi0xtgen}) is still a three-parameter family of states, so in the sections below we consider some special parameter choices to illustrate the role of each parameter. 

\subsubsection{Special case: Mixing identical anharmonic oscillators}
Consider first the case of two identical oscillators, $c_1 = c_2 \equiv c$. (Actually, this choice corresponds to a larger family in the original parameterization, since we require only $b_1/\sqrt{a_1} = b_2/\sqrt{a_2}\equiv c$. Note this implies ${\rm sign}(b_1)={\rm sign}(b_2)$, since the $a_i$ are always positive.) We further choose the special mixing angle $\theta = \pi/4$. This allows for the cancellation of some terms from Eq.(\ref{eq:psi0xtgen}) which otherwise lead, in the process of calculating reduced density matrices, to more complicated integrals with analytic results known only in the form of infinite series. With $\theta = \pi/4$, the ground state wave function is now particularly simple, given by 
\begin{equation}
\label{psitpi4}
     \psi_{0}(x_1,x_2, c, \pi/4) =  A^2 \exp\left[-\frac{c}{2}\left(x_1^2+x_2^2\right) -\frac{1}{8}\left(x_1^4+6x_1^2x_2^2+x_2^4\right)\right]\,. 
\end{equation}
The reduced density matrix, $\rho(x_1, x_1^\prime) = \int dx_2 \psi_0(x_1, x_2)\psi_0^*(x_1^{\prime}, x_2)$ is
\begin{equation}
\label{eq:rhoIdentPiover4}
   \rho(x_1,x_1^{\prime},c, \pi/4)= \frac{1}{2}A^4
    \exp\left[-\frac{c}{2}(x_1^2+x_1^{\prime2})-\frac{1}{8}
    (x_1^4+x_1^{\prime4})\right]f(x_1, x_1^\prime, c) \,,
\end{equation}
with normalization factor $A^4$ given by Equation \ref{eq:At4norm} and the function $f(x_1, x_1^\prime, c)$ defined in terms of $u(x_1,x_1^\prime, c) = 4c+3(x_1^2 + x_1^{\prime 2})$ as
\begin{align}
    f(x_1, x_1^\prime, c) =
\begin{dcases}
    \sqrt{u}
    \exp\left[\frac{u^2}{32}\right]K_{\frac{1}{4}}\left[\frac{u^2}{32}\right]  & \text{for} \,\, x_1^2+x_1^{\prime2} > -\frac{4c}{3} 
    \\
    \sqrt{\frac{1}{2}}\,\Gamma\left(\frac{1}{4}\right) & \text{for} \,\, x_1^2+x_1^{\prime2} = -\frac{4c}{3}
    \\
   \pi\sqrt{\frac{-u}{2}}
    \exp\left[\frac{u^2}{32}\right]\left(I_{-\frac{1}{4}}\left[\frac{u^2}{32}\right] + I_{\frac{1}{4}}\left[\frac{u^2}{32}\right] \right) & \text{for} \,\, x_1^2+x_1^{\prime2} < -\frac{4c}{3}\,.
\end{dcases}
\end{align}
Here $I_{\pm\frac{1}{4}}[z]$ and $K_{\frac{1}{4}}[z]$ are modified Bessel functions of the first and second kind, respectively. This function is continuous across $x_1^2+x_1^{\prime2} = -4c/3$. When $c>0$, the definition simplifies since in this case it is always true that $ x_1^2+x_1^{\prime2} > -4c/3$. 

\begin{figure}[htbp]
    \centering
    \includegraphics[width=\textwidth]{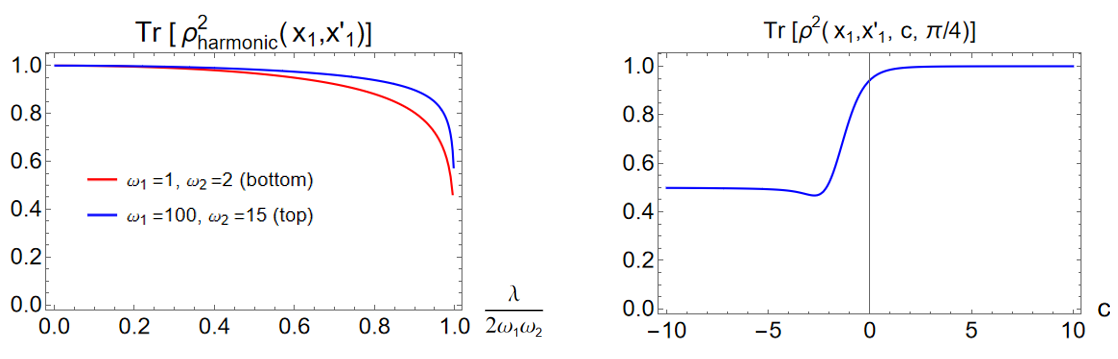}
    \caption{\textbf{Left:} Purity of the reduced density matrix for $x_1$ for two linearly coupled harmonic oscillators, shown for two different sets of values $\omega_1, \omega_2$. In both cases the purity stays relatively high until it plummets near the maximum allowed coupling strength $\lambda = 2\omega_1\omega_2$. \textbf{Right:} The purity of the reduced state $\rho(x_1,x_1^{\prime},c)$ as a function of $c$ for two identical anharmonic oscillators mixed at the special angle $\pi/4$. The constant $c$ is the only parameter in all interaction and quadratic terms in that case. In the general case of non-identical oscillators mixed at arbitrary angles (not shown), it is possible to obtain subsystem purities significantly less than 1/2. \label{fig:purityfuncofc2idpiov4}}
\end{figure}

The variance of $x_1$ after tracing out $x_2$ is always exactly equal to the variance of an individual oscillator with the same $c$ value. (Actually, this remains true for two identical oscillators coupled via any mixing angle, discussed below.) The purity of the reduced state $\rho(x_1,x_1^{\prime},c, \pi/4)$ is shown as a function of $c$ in Figure \ref{fig:purityfuncofc2idpiov4}. It approaches 1 as $c \rightarrow + \infty$ and $1/2$ as $c \rightarrow -\infty$. The limit of 1/2 for large negative $c$ is approached from below after a modest initial dip to about 0.47 near $c= -2.7$, slightly beyond the $c = -\sqrt{3}$ boundary between the double and triple-well potentials. The results for $\theta = \pi/4$ do not provide strict lower bounds as compared to results for other angles at fixed $c$ but do serve as a useful reference point for the general behavior. For two identical oscillators, subsystem purity at some $c$-dependent subset of other mixing angles may dip slightly, but typically not drastically, below that found at $\pi/4$. Beyond this special case, the subsystem purity can be significantly lower than $1/2$  for two non-identical oscillators and some (possibly large) fraction of mixing angles, especially when at least one of the $c$ values corresponds to a strongly non-Gaussian state in the decoupling frame. 

\begin{figure}[htbp]
\begin{centering}
  \includegraphics[width=.4\textwidth]{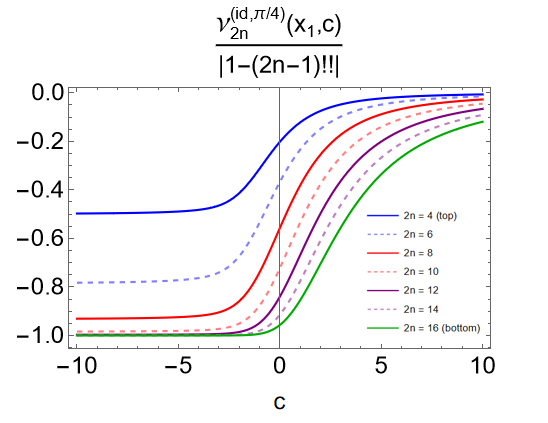}
  \includegraphics[width=.4\textwidth]{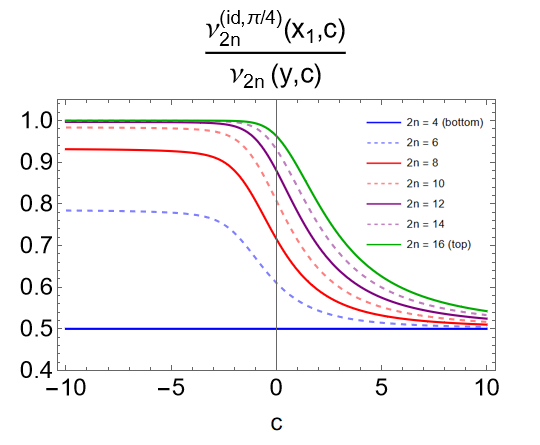}
    \caption{\textbf{Left}: Combined plots of the moments, $\nu_{2n}^{(\text{id},\pi/4)}(x_1,c)/|1-(2n-1)!!|$, of one of a pair of identical anharmonic oscillators mixed at angle $\theta=\pi/4$, after tracing out the other, for $2n = 4, 6, ... , 16$. \textbf{Right:} Ratios (as a function of $c$) of $2n^{th}$ moments of $x_1$ (after trace out of $x_2$) divided by the corresponding $2n^{th}$ moments of the individual oscillators with same $c$. \label{fig:nucratiofuncs}}
\end{centering}
\end{figure}

The primary qualitative result for mixing two identical oscillators is that quadrature measurements are always more Gaussian than the single-oscillator statistics. However, higher order moments inherit more of the non-Gaussianity of the single-oscillator system than lower-order moments do. For higher even moments, results are displayed in Figure \ref{fig:nucratiofuncs}. We show results for mixing the oscillators at $\theta = \pi/4$, which gives the most weakly non-Gaussian single-oscillator reduced state possible for any value of $c$. At this angle, the kurtosis of the reduced state is always exactly halved compared to the single oscillator, for all values of $c$ (the bottom solid blue line in Figure \ref{fig:nucratiofuncs}). The higher even moments also approach half of the single oscillator values at large positive $c$. However, at large negative $c$, higher moments remain very non-Gaussian and as $2n$ becomes large, the ratio of mixed state to pure state non-Gaussianity approaches 1. 

\begin{figure}[htbp]
\begin{centering}
  \includegraphics[width=.4\textwidth]{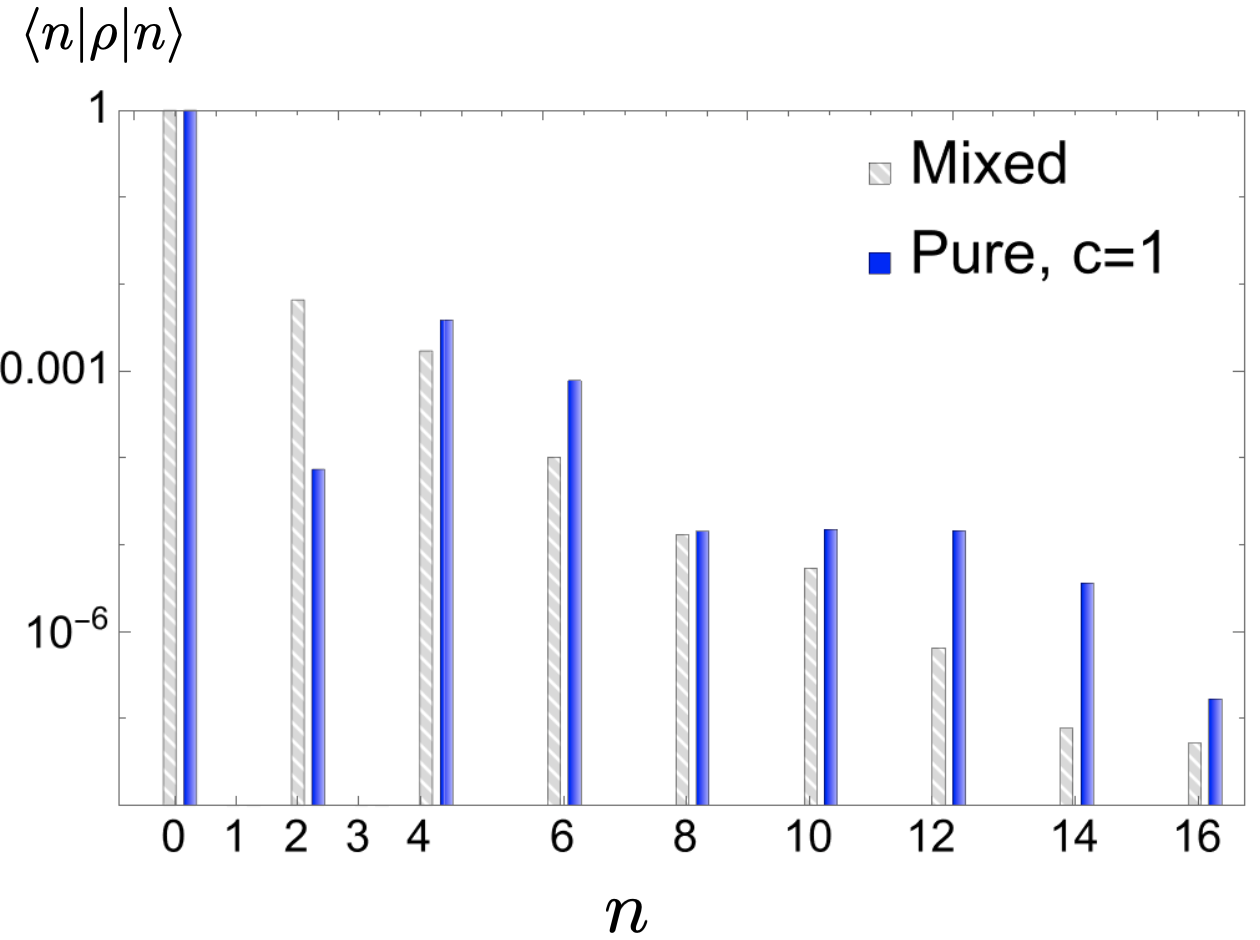}
  \includegraphics[width=.4\textwidth]{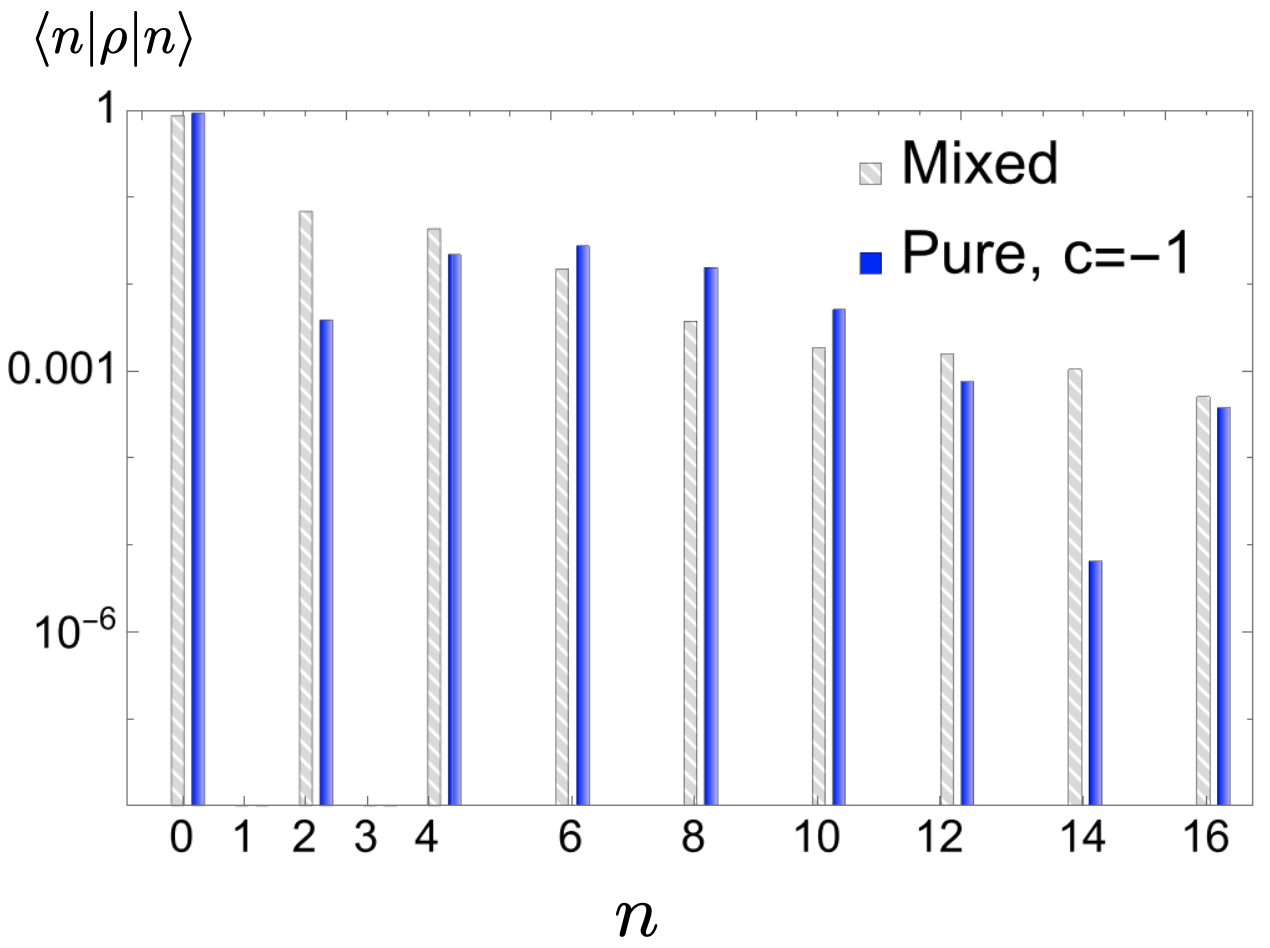}
    \caption{The number operator statistics for the mixed state (gray hatched) after tracing out one of an identical pair of anharmonic oscillators, mixed at angle $\theta = \pi/4$, compared to the single oscillator statistics (solid blue), for $c=1$ (left) and $c=-1$ (right). \label{fig:MixedAnharmonicNpops}}
\end{centering}
\end{figure}

The number operator statistics are shown in Figure \ref{fig:MixedAnharmonicNpops}, again comparing the single-oscillator statistics to those of the reduced state. To compare most closely to the thermal result for linearly coupled harmonic oscillators, we have again used the variance to define the frequency for the Fock states, $\langle x_1^2\rangle\equiv \frac{1}{2\Omega_T}$. Since the mixing is particularly simple here, this variance is also that of the individual oscillators, which makes the comparison to the pure anharmonic state straightforward. As the figure shows, the mixed state loses the oscillatory behavior seen in the single oscillator case. Only even number states have non-zero occupation, but the overall trend is a simple fall-off with $n$.

\begin{figure}
    \centering
    \includegraphics[width=.9\textwidth]{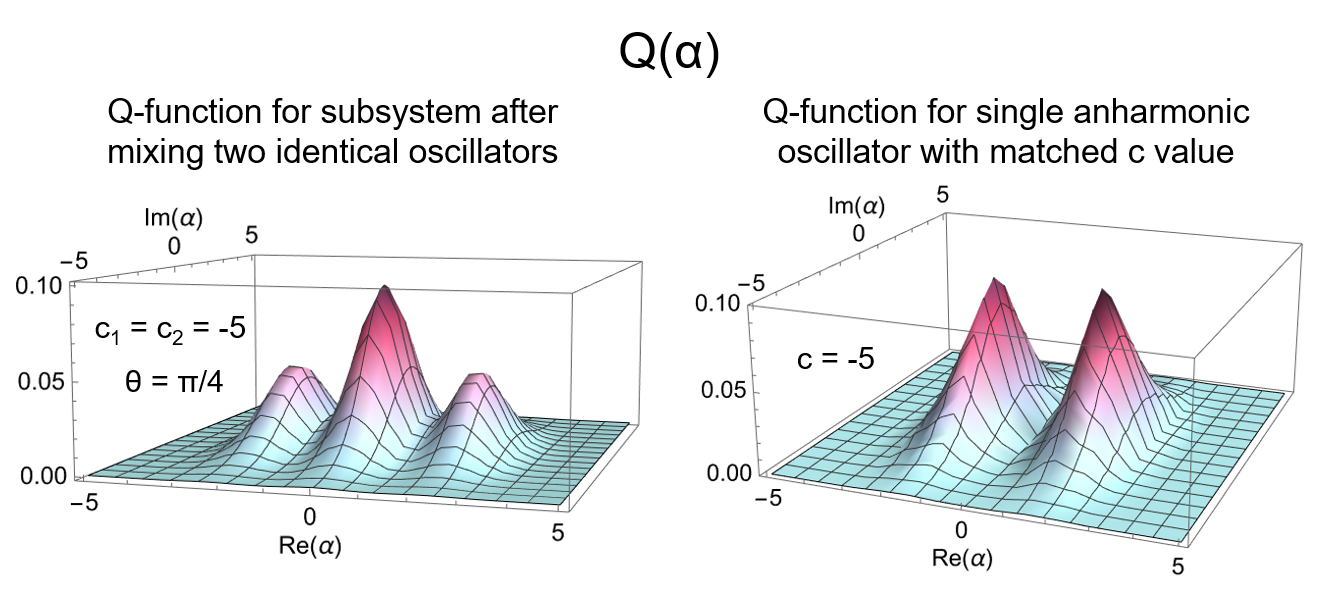}
    \caption{Left: Mixed-state Q-function for the reduced density matrix obtained from two identical $c=-5$ oscillators coupled via mixing angle $\theta = \pi/4$. Right: Pure state Q-function for single $c=-5$ oscillator. The mixed state is more weakly non-Gaussian with its centralized hump and smaller outer humps as compared pure state with matched $c$ value.  \label{fig:Qfunc2idpiov4vs1Dc-5} }
\end{figure}

Finally, the Q-functions of the single-oscillator reduced states similarly show a return toward Gaussianity. For example, Figure \ref{fig:Qfunc2idpiov4vs1Dc-5} shows the Q-function of an isolated sextic oscillator is shown alongside that of the mixed state reduced density matrix obtained from two identical oscillators mixed at $\theta=\pi/4$, for  $c=-5$. The mixed state appears to be more Gaussian than the pure state, as the central hump has reappeared while the two outer humps have shrunk in comparison.

For these mixed states, the stellar rank $r_{*}$ can be formally defined by looking at all possible state-decompositions of the density matrix, $r_{*}(\rho)={\rm inf}_{\{p_i,\psi_i\}}{\rm sup}\,r_{*}(\psi_i)$, for all possible $\rho=\sum_i p_i|\psi_i\rangle\langle\psi_i|$ \cite{Chabaud:2020}. In practice, the degree of quantum non-Gaussianity can also be bounded by comparing the fidelity with a target state of know stellar rank $k$ to the separation between states of rank $k$ and those of rank $k-1$ \cite{Chabaud:2020,Chabaud:2021}. (However, it cannot be used to certify infinite stellar rank since they are not robustly separated; that is, they can be arbitrarily well-approximated by states of finite stellar rank.) This calculation may be easier than actually computing the stellar rank, but it is still non-trivial. A target witness state must be chosen, and then the threshold bound on fidelity with the target state that signals non-Gaussianity must be found. Numerical work useful for several non-Gaussian witnesses of low stellar rank was done in \cite{Chabaud:2021pnh, Fiurasek:2022utu}, but those witnesses are not able to certify the non-Gaussianity of either the pure anharmonic oscillator states, or the resulting mixed states. This can be seen by from the small probability of finding $\langle \hat{N}\rangle=2$ and higher from Figure \ref{fig:MixedAnharmonicNpops} and comparing to Table 1 or the appendix of \cite{Chabaud:2021pnh}. While the statistics can be shifted by choosing a different frequency, we did not find a prescription for choosing a frequency that allows a detection of non-Gaussianity with the witnesses from \cite{Chabaud:2021pnh}. It would be interesting to construct a witness more suited to the class of non-Gaussian states suggested by the QES anharmonic oscillator. 

\subsubsection{General case: Mixing non-identical anharmonic oscillators}
\label{sec:generalcoupledanharm}
\begin{figure}[hbtp]
    \centering
    \includegraphics[width=.4\textwidth]{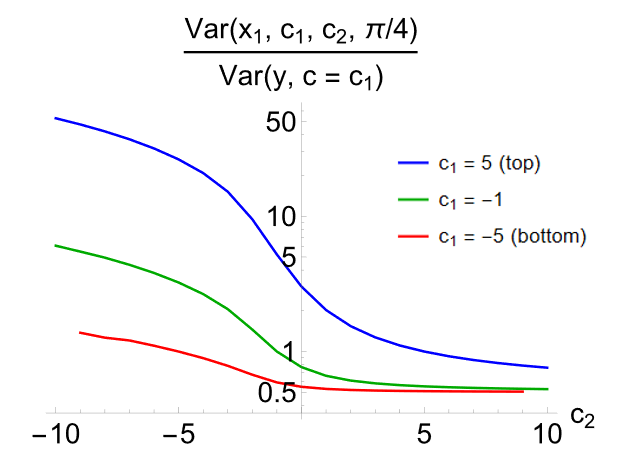}
    \includegraphics[width=.4\textwidth]{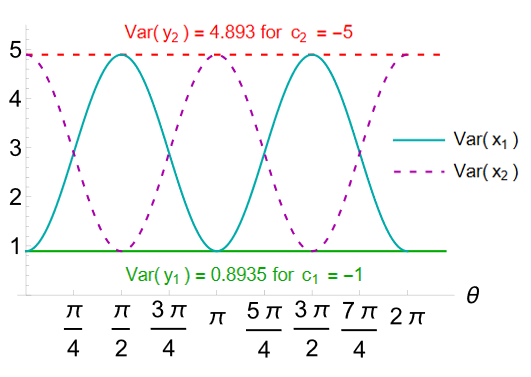}
    \caption{Left: Ratio of variance of $x_1$ for coupled non-identical oscillators as a function of $c_2$, shown for three fixed values of $c_1$ and mixing angle $\pi/4$.  Right: An example of variance of $x_1$ and $x_2$ as a function of the mixing angle $\theta$ after tracing out the other oscillator, for non-identical oscillators with $c_1 = -1$ and $c_2 = -5$. Variance behavior for other pairs of $c_1$, $c_2$ values is qualitatively the same, following Eq.(\ref{eq:variancex1x2}).
    \label{fig:varx1nonid}}
\end{figure}

Away from the limit of identical oscillators, it is challenging to obtain useful analytic expressions for the single-oscillator reduced density function (although an infinite series form can be written). The mixed-state variance, however, can be simply expressed in terms of the single oscillator values. When a $c = c_1$ oscillator in variable $y_1$ is mixed with a $c = c_2$ oscillator in variable $y_2$ at angle $\theta$, the variances satisfy $\text{Var}(x_1) +  \text{Var}(x_2) =  \text{Var}(y_1) +  \text{Var}(y_2)$, with each individual piece given by
\begin{align}
\label{eq:variancex1x2}
    \text{Var}(x_1) =& |\text{Var}(y_2) - \text{Var}(y_1)|\sin^2(\theta) + \min\{\text{Var}(y_1), \text{Var}(y_2)\}\\\nonumber
    \text{Var}(x_2) = &|\text{Var}(y_2) - \text{Var}(y_1)|\cos^2(\theta) + \min\{\text{Var}(y_1), \text{Var}(y_2)\}\,,
\end{align}
at least to a very good approximation. Figure \ref{fig:varx1nonid} shows that that the coupling with a second oscillator can cause the variance of $x_1$ to become arbitrarily larger than the variance of an individual oscillator with $c=c_1$, with increasingly strong effect as $c_2\rightarrow -\infty$ and/or $c_1 \rightarrow + \infty$. The similar increase in variance that occurs in the case of linearly coupled harmonic oscillators was emphasized in \cite{Balasubramanian2021}. On the other hand, it is also possible for coupling to cause the variance of $x_1$ to be significantly smaller than that of the corresponding isolated oscillator. This occurs roughly for $c_2$ somewhat greater than $c_1$, since from Eq.(\ref{eq:variancex1x2}), $\text{Var}(x_1)<\text{Var}(y_1)$ requires mixing with an oscillator with smaller variance, $\text{Var}(y_2)<\text{Var}(y_1)$.
Beyond the variance, we use a few fully numerical examples to illustrate some general features of this larger class of non-Gaussian mixed states. Non-exact analytic expressions can provide a good approximation within a limited range of the parameter space (see Appendix \ref{app:nonid}), but they are not always valid. 
 
\begin{figure}[hbtp!]
    \centering
    \includegraphics[width=.9\textwidth]{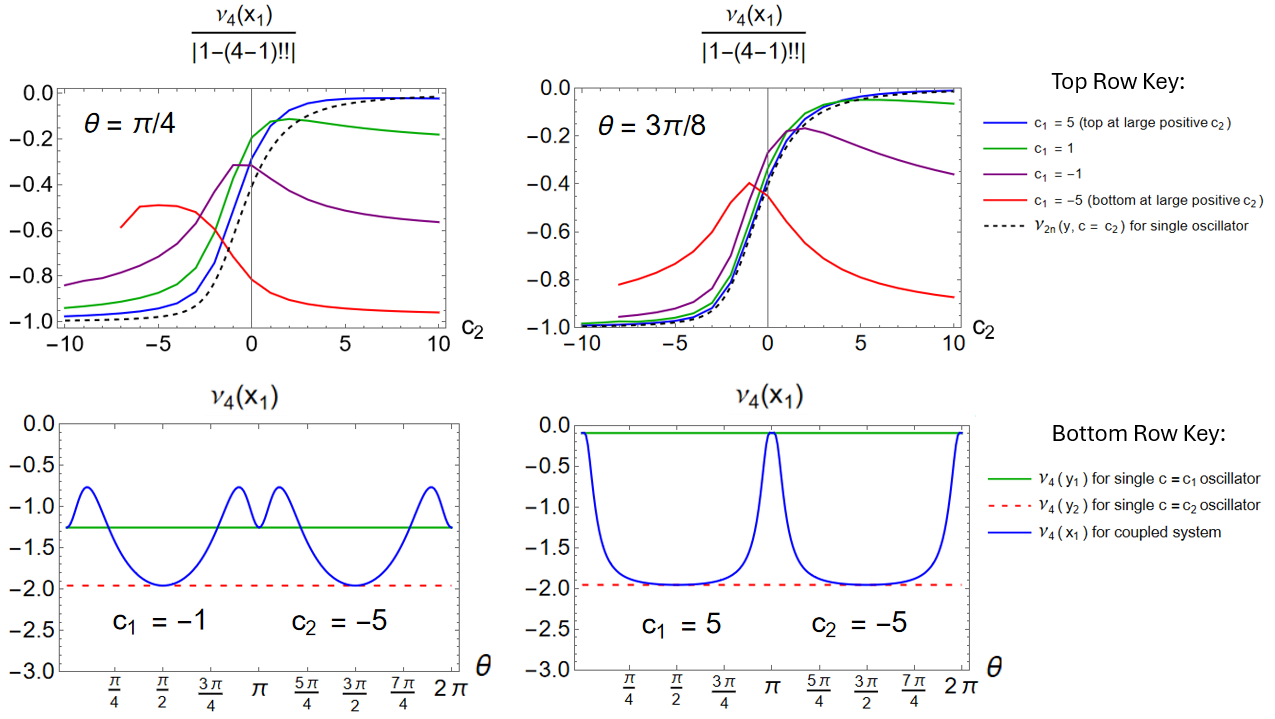}
    \caption{\textbf{Top Row:} Subsystem moment $\nu_4(x_1)$, scaled by the maximum value $|\nu_{4,{\rm max}}|$, as a function of $c_2$ at fixed mixing angle $\theta$, plotted for several values of $c_1$. \textbf{Bottom Row:} Subsystem moment $\nu_4(x_1)$ as a function of $\theta$, plotted for fixed pairs of values $c_1, c_2$. For comparison, flat solid green and and dotted red lines show the corresponding values of the excess fourth moment of a single uncoupled oscillator with $c = c_1$ and $c = c_2$, respectively.}
    \label{fig:nonidKurtosis}
\end{figure}

The most important qualitative trend for non-identical oscillators is that while in general the reduced-state oscillator is more Gaussian than only the most non-Gaussian of the parents, there are some parts of parameter space where it is more Gaussian than both parents. This implies that, for fixed $c_1$ of the measured oscillator, the state after mixing with another oscillator may be either more or less Gaussian than the state of a single oscillator. The reduced state moments still follow the same pattern observed for single oscillators, where higher moments are relatively more non-Gaussian than lower moments. It is therefore sufficient to compare a single higher order term for the mixed states to that of the single-oscillator pure states. We use kurtosis in Figure \ref{fig:nonidKurtosis} for this purpose. 
The bottom left panel illustrates how the relative Gaussianity depends on mixing angle, and the bottom right panel shows that not all pairs of oscillators have a range of mixing angles for which the coupled oscillator statistics are significantly more Gaussian than either of the parents. Although we leave a more detailed study of the Q-functions and number operator results for future work, we expect that the trends observed in the single oscillator and identically coupled oscillators that relate those measures to the features of the $\hat{x}$ statistics will carry over.

\section{Conclusions}
\label{sec:conclude}
There are a number of ways that unobserved degrees of freedom may affect the dynamics of those observed \cite{burgess2020introduction}. The most familiar treatment in particle physics, and most relevant for collider physics, is via low-energy effective Lagrangians that incorporate the effects of degrees of freedom that are heavy compared to the scale probed. On the other hand, if the unobserved degrees of freedom are light and dynamical, a full open-system treatment may be required. Such situations are nearly inevitable in laboratory quantum systems \cite{BreuerBook,Rotter_2015}, but are also increasingly studied in cosmology \cite{Agon:2014uxa,Boyanovsky:2015tba,Shandera:2017qkg,Boyanovsky:2018fxl,Mirbabayi:2020vyt,Jana:2020vyx,Zarei:2021dpb,Hsiang:2021kgh,Burgess:2022rdo,Cao:2022kjn, Colas:2024xjy,Bhattacharyya:2024duw,Burgess:2024eng,Bowen:2024emo,Salcedo:2024smn}. In this paper, we consider an intermediate case, where the extra degrees of freedom appear in the effective dynamics as non-dynamical coupling constants, but the coupling constants vary, taking a range of values inherited from the complexity of the quantum state of the environment at the time the couplings were determined. 

We proposed several distributions that may be used for studying this kind of disorder, with parameters drawn either from the family of non-Gaussian continuous distributions formed by quadrature measurements of $\rho(x,x^{\prime})$ as given by Eq.(\ref{eq:rhoIdentPiover4}), or distributions of the discrete values (number statistics) drawn from the histograms like those in Figure \ref{fig:MixedAnharmonicNpops}. Specifically, we have considered environments composed of two continuous variable systems, in global ground states that ranged from weakly to strongly non-Gaussian in terms of quadrature measurements. The division into the measured degree of freedom and the unmeasured variables was controlled by a parameter ($\theta$ in Eq.(\ref{eq:rotdef})), where for $\theta=0$ the division is into two non-interacting degrees of freedom. Otherwise, the interactions contain all symmetric terms up to order six, all of similar strength. Some features of these states are restricted by the requirement of exactly solvable ground states, while others may be more general. In general, the statistics of these states provide an example of a class of non-Gaussian distributions for disorder that are natural (eg, generated by nonlinear Hamiltonians) and that are fully characterized beyond perturbation theory. Furthermore, we clarified how features of the disorder distribution contain signatures of the nature of the environmental fields, and how the features may be distorted from the conclusions one would assuming a single environmental field (Figures \ref{fig:nucratiofuncs}, \ref{fig:MixedAnharmonicNpops}, and \ref{fig:nonidKurtosis}).

In a laboratory context, this phenomenon is disorder from measurement. It may be engineered to study systems with disorder beyond the commonly studied flat and Gaussian distributions, but it is also a plausible phenomenon for small-scale, naturally occurring systems with a dynamical response to an environment. A separation of scales is needed between system and environment in order for the environmental degrees of freedom to provide effective, locally fixed, coupling constants for the dynamics of the system. Although the specific interaction terms we considered are from a particular class (that is, they appear fine-tuned in the sense of Eq.(\ref{eq:6generic})-Eq.(\ref{eq:lambdaConstrained}) and they are not connected to free, Gaussian fields by a smooth limit), they may serve as fully-defined non-Gaussian examples with which to identify new phenomena or signatures in systems with this kind of disorder. While understanding the effects of disorder distributions in spin systems is numerically challenging, recent advances \cite{Rigol:2006} tested on multi-modal distributions \cite{Tang:2015, Mulanix:2019, Park:2021, Abdelshafy:2024mnp} may allow computations for an even wider range of scenarios. Such laboratory studies could uncover specific nonlinear phenomena that can serve as a guideline for novel cosmological signatures of physics beyond the Standard Model. It would be interesting to work out a concrete dynamical scenario in the context of cosmology that leads directly to such disorder, perhaps along this lines of a generalized treatment of moduli stabilization appropriate when a hidden sector contains light fields \cite{Balasubramanian2021}. Our work here provides a set of possible implications for such a scenario, calculable beyond the specialized case of only linear couplings and Gaussian states. 

\acknowledgments
We thank Archana Kamal for helpful discussions. This work was supported by the National Science Foundation under PHY-1719991 and PHY-2310662.
\pagebreak

\appendix 
\section{Analytic expressions for 1D anharmonic oscillators}
\label{app:analytic}
The normalization constant $A$ for the ground state wave functions of a single anharmonic oscillator, Eq.(\ref{eq:anhamronicWF}), is given by
\begin{align}
\label{eq:Atnorm}
A = a^{-\frac{1}{8}}\tilde{A} =
\begin{dcases}
\left(\frac{2}{c}\right)^\frac{1}{4}\frac{1}{\sqrt{e^{\frac{1}{4}c^2}K_\frac{1}{4}\left(\frac{c^2}{4}\right)}}\, & \text{if} \,\,c > 0
\\
\left(-\frac{4}{\pi^2c}\right)^{\frac{1}{4}}\frac{1}{\sqrt{e^{\frac{1}{4}c^2}\left(I_{-\frac{1}{4}}\left[\frac{c^2}{4}\right]+I_{\frac{1}{4}}\left[\frac{c^2}{4}\right]\right)}}\, & \text{if}\,\,c<0 
\\
\frac{2^\frac{3}{8}}{\sqrt{\Gamma\left[\frac{1}{4}\right]}}\, & \text{if}\,\,c = 0\,,
\end{dcases}
\end{align}
where $I_{\nu}(x)$ and $K_{\nu}(x)$ are the modified Bessel functions of the second kind, and $\Gamma(z)$ is the usual gamma function.

For $c$ an arbitrary real number and $n$ any positive integer, the $2n^{th}$ raw even moment with respect to the rescaled variable $y$ is
\begin{equation}
\expval{y^{2n}} =
\begin{dcases}
     \frac{\Gamma\left[\frac{1}{2}+n\right]U\left[\frac{1}{4}+\frac{n}{2}, \frac{1}{2}, \frac{c^2}{2}\right]}{2^{\frac{n}{2}-\frac{1}{4}}\sqrt{c}\,\exp\left[\frac{c^2}{4}\right] K_{\frac{1}{4}}\left[\frac{c^2}{4}\right]}\,, & \text{if}\,\, c > 0
     \\
      \frac{2^{\frac{n}{2}+\frac{1}{4}}\left(\Gamma\left[\frac{1}{4}+\frac{n}{2}\right]\,_1F_1\left[\frac{1}{4}+\frac{n}{2}, \frac{1}{2}, \frac{c^2}{2}\right] - \sqrt{2}\,c\,\Gamma\left[\frac{3}{4} + \frac{n}{2}\right] \,_1F_1\left[\frac{3}{4}+\frac{n}{2}, \frac{3}{2}, \frac{c^2}{2}\right]\right)}{\pi\sqrt{-c}\,\exp\left[\frac{c^2}{4}\right] \left(I_{-\frac{1}{4}}\left[\frac{c^2}{4}\right]+I_{\frac{1}{4}}\left[\frac{c^2}{4}\right]\right)}\,, & \text{if}\,\, c <0\\
      \frac{2^\frac{n}{2} \Gamma\left[\frac{1}{4}+\frac{n}{2}\right]}{\Gamma\left[\frac{1}{4}\right]}\,, & \text{if}\,\, c = 0\,,
\end{dcases}
\end{equation}
where $_1F_1[a,b,c]$ is the Kummer confluent hypergeomentric function and $U[a,b,c]$ is the Tricomi confluent hypergeometric function which can be expressed as
\begin{equation}
U[a,b,c] = \frac{1}{\Gamma[a]} \int_0^\infty dt\, e^{-ct}t^{a-1}(1+t)^{b-a-1}\,.
\end{equation}
Noting $\expval{y}=0$, the variance is thus
\begin{align}
\label{varytilde}
\text{Var}(y,c) = \expval{y^2} =  
\begin{dcases}
\frac{\Gamma\left[\frac{3}{2}\right]U\left[\frac{3}{4}, \frac{1}{2}, \frac{c^2}{2}\right]}{2^{\frac{1}{4}}\sqrt{c}\,\exp\left[\frac{c^2}{4}\right] K_{\frac{1}{4}}\left[\frac{c^2}{4}\right]}\,, & \text{if} \,\, c>0
\\
 \frac{2^\frac{3}{4}\left(\Gamma\left[\frac{3}{4}\right]\,_1F_1\left[\frac{3}{4}, \frac{1}{2}, \frac{c^2}{2}\right] - \sqrt{2}\,c\,\Gamma\left[\frac{5}{4}\right]\,_1F_1\left[\frac{5}{4}, \frac{3}{2}, \frac{c^2}{2}\right]\right)}{\pi\sqrt{-c}\,\exp\left[\frac{c^2}{4}\right] \left(I_{-\frac{1}{4}}\left[\frac{c^2}{4}\right]+I_{\frac{1}{4}}\left[\frac{c^2}{4}\right]\right)}\,, & \text{if}\,\, c <0
\\
\frac{\sqrt{2} \Gamma\left[\frac{3}{4}\right]}{\Gamma\left[\frac{1}{4}\right]}\,, & \text{if} \,\, c=0\,.
\end{dcases}
\end{align}

The $2n^{th}$ even excess moment can be written as
\begin{multline}
    \nu_{2n}(y, c) = \frac{\expval{y^{2n}}}{(\text{Var}(y,c))^n} - (2n-1)!! = \\
    \\
    \begin{dcases}
    \frac{B_{+}(n,c)\Gamma\left[\frac{1}{2}+n\right]U\left[\frac{1}{4}+\frac{n}{2}, \frac{1}{2}, \frac{c^2}{2}\right]}{\left(\Gamma\left[\frac{3}{2}\right]U\left[\frac{3}{4}, \frac{1}{2}, \frac{c^2}{2}\right]\right)^n} -(2n-1)!!\,, & \text{if} \,\, c>0
     \\
     \\
       \frac{B_{-}(n,c)\pi^{n-1}\left(\Gamma\left[\frac{1}{4}+\frac{n}{2}\right]\,_1F_1\left[\frac{1}{4}+\frac{n}{2}, \frac{1}{2}, \frac{c^2}{2}\right] - \sqrt{2}\,c\,\Gamma\left[\frac{3}{4} + \frac{n}{2}\right]\,_1F_1\left[\frac{3}{4}+\frac{n}{2}, \frac{3}{2}, \frac{c^2}{2}\right]\right)}{\left(\Gamma\left[\frac{3}{4}\right]\,_1F_1\left[\frac{3}{4}, \frac{1}{2}, \frac{c^2}{2}\right] - \sqrt{2}\,c\,\Gamma\left[\frac{5}{4}\right]\,_1F_1\left[\frac{5}{4}, \frac{3}{2}, \frac{c^2}{2}\right]\right)^n} 
       \\
       \hspace{12cm}- (2n-1)!!\,,
     & \text{if} \,\, c<0
    \\
   \frac{\Gamma\left[\frac{1}{4}+\frac{n}{2}\right]\left(\Gamma\left[\frac{1}{4}\right]\right)^{n-1}}{\left(\Gamma\left[\frac{3}{4}\right]\right)^n} - (2n-1)!!\,, & \text{if} \,\, c=0\,,
    \end{dcases}
\end{multline}
where
\begin{equation}
    B_{\pm}(n,c) \equiv \left(\frac{\sqrt{\pm c}\,\exp\left[\frac{c^2}{4}\right]K_{\frac{1}{4}}\left[\frac{c^2}{4}\right]}{2^{\frac{1}{4}}}\right)^{n-1}\,\,.
\end{equation}

\section{Details of coupled systems}
\label{app:analyticcoupled}
In the main text we constructed coupled anharmonic oscillator examples by assuming the existence of a decoupling frame. Here we show how to see where these results apply from the more traditional perspective, where the definitions of the free fields are handed down first, from some encompassing theory, and then couplings are introduced. In that case, the quasi-exactly solvable family of sextic oscillators is related to a class of Hamiltonians with even terms up to sixth order\footnote{Interestingly for the story of moduli stabilization and disorder, potentials with terms only up to sixth order may arise naturally for fields that descend from the NS-NS two-form potential B in Type II string theory \cite{ Polchinski_1998, McAllister:2014mpa}. We thank Timm Wrase for pointing this out to us.}, 
\begin{align}
\label{eq:6generic}
H =& \frac{1}{2}(p_{x_1}^2 + p_{x_2}^2) +\frac{1}{2}(\omega_1^2x_1^2+\omega_2^2x_2^2)+\sum_{k=1}^{3}\left(\sum_{i,j|i+j=2k} \lambda_{ij}x_1^ix_2^j\right)  - B_0 \,,
\end{align}
with constrained frequencies $\omega$ and couplings $\lambda_{ij}$. The form of the constraints can be expressed by first rewriting the Hamiltonian as
\begin{align}
\label{eq:coupledAnharmonic}
    H =& \frac{1}{2}(p_{x_1}^2 + p_{x_2}^2) + (\vec{\lambda}_{02}\cdot\vec{X}_{02}+ \vec{\lambda}_{04}\cdot\vec{X}_{04}+\vec{\lambda}_{06}\cdot\vec{X}_{06})\\\nonumber
    &+\lambda_{11} x_1x_2 + \lambda_{22}x_1^2x_2^2 +\lambda_{33}x_1^3x_2^3\\\nonumber
    &+\vec{\lambda}_{13}\cdot\vec{X}_{13}+\vec{\lambda}_{15}\cdot\vec{X_{15}}+\vec{\lambda}_{24}\cdot\vec{X_{24}}\,,
\end{align}
where, for example,
\begin{equation}
    \vec{\lambda}_{13}\cdot\vec{X}_{13}=\lambda_{13,1}x_1x_2^3+\lambda_{13,2}x_1^3x_2\,.
\end{equation}
Then within each $\vec{\lambda}_{ij}$ the two components are related by
\begin{equation}
\label{eq:lambdaConstrained}
    \left(\begin{array}{c}
    {\lambda}_{ij,1}\\
    {\lambda}_{ij,2}\\
    \end{array}
    \right)=\left(\begin{array}{cc}
    p&1-p\\
    1-p&p\\
    \end{array}\right)\left(\begin{array}{c}
    {\alpha}_{ij}\\
    {\beta}_{ij}\\
    \end{array}
    \right)\,,
\end{equation}
where $p=\cos^2(\theta)$. Furthermore, the $\alpha_{ij}$ and $\beta_{ij}$ are functions the five parameters appearing in the uncoupled potentials in Eq.(\ref{eq:potential}), $a_1$, $a_2$, $b_1$, $b_2$, $n$ (where of course the difference between $\tilde{y}$ or $y$ notation is unimportant). Tracing out one of the oscillators in an $\theta\neq0$ frame generically results in a non-Gaussian mixed state for the other oscillator, which can be found exactly. 

\subsection{Mixing anharmonic oscillators}
\label{sec:exact}
The most general ground state wave function for the system of coupled, quasi-exactly solvable sextic oscillators is 
\begin{multline}
\label{psi0xgen}
   \tilde{\psi}_0(\tilde{x}_1, \tilde{x}_2)= \tilde{A}_1\tilde{A}_2\exp{-\frac{b_1}{2}\left[(\cos\theta) \tilde{x}_1 - (\sin\theta) \tilde{x}_2\right]^2
    -\frac{b_2}{2}\left[(\sin\theta) \tilde{x}_1 + (\cos\theta)\tilde{x}_2\right]^2}\\
   \times \exp{-\frac{a_1}{4}\left[(\cos\theta) \tilde{x}_1 - (\sin\theta) \tilde{x}_2\right]^4 -\frac{a_2}{4}\left[(\sin\theta) \tilde{x}_1 + (\cos\theta) \tilde{x}_2\right]^4}\,.
\end{multline}

\subsubsection{Special case: Identical oscillators coupled via mixing  angle \texorpdfstring{{$\pi/4$}}{Pi/4}}
\label{sec:identicalpi4}
Consider first two identical oscillators in the uncoupled frame with $n=0$, $a_1=a_2\equiv a$, $b_1=b_2\equiv b$. Choosing the special mixing angle $\theta = \pi/4$ so that $\sin\theta = \cos\theta = 1/\sqrt{2}$ yields 
\begin{align}
\label{Hpi4}
     \tilde{H}_{\frac{\pi}{4}}^{(id)}(\tilde{x}_1,\tilde{x}_2) =  &\frac{1}{2}(p_{\tilde{x}_1}^2 + p_{\tilde{x}_2}^2)
     + (b^2-3a)(\tilde{x}_1^2 + \tilde{x}_2^2)
      + ab(\tilde{x}_1^4+\tilde{x}_2^4)\color{black} + \frac{1}{4}a^2(\tilde{x}_1^6 + \tilde{x}_2^6) \color{black}\\\nonumber
&     + \frac{15}{4}a^2(\tilde{x}_1^4\tilde{x}_2^2 + \tilde{x}_1^2\tilde{x}_2^4)\color{black} 
    + 6ab(\tilde{x}_1^2\tilde{x}_2^2)\color{black}
    +2b\,,
\end{align}
with ground state wave function
\begin{equation}
\label{psipi4}
     \tilde{\psi}_{0,\frac{\pi}{4}}^{(id)}(\tilde{x}_1,\tilde{x}_2) =  A^2 \exp{-\frac{b}{2}\left(\tilde{x}_1^2+\tilde{x}_2^2\right) -\frac{a}{8}\left(\tilde{x}_1^4+6\tilde{x}_1^2\tilde{x}_2^2+\tilde{x}_2^4\right)}\,.
\end{equation}
For both oscillators in the ground state and $x_2$ traced out in the coupled frame, the reduced density matrix for $x_1$ can be written as 
\begin{equation}
\label{eq:rhoredab}
    \tilde{\rho}(\tilde{x}_1,\tilde{x}_1^{\prime}, a,b)= \frac{1}{2}A^4
    \exp\left[-\frac{b}{2}(\tilde{x}_1^2+\tilde{x}_1^{\prime2})-\frac{a}{8}
    (\tilde{x}_1^4+\tilde{x}_1^{\prime4})\right]\tilde{f}(\tilde{x}_1, \tilde{x}_1^\prime, a,b) \,,
\end{equation}
The reduced density matrix $\rho(x_1, x_1^\prime) = \int dx_2 \psi_0(x_1, x_2)\psi_0^*(x_1^{\prime}, x_2)$ in the rescaled coordinates is simply related to that for the original parameters by $\rho(x_1, x_1^\prime,c) = a^{-\frac{1}{4}} \tilde{\rho}(\tilde{x}_1,\tilde{x}_1^{\prime}, a, b)$.

The normalization $\tilde{A}^4$ is given by
\begin{align}
\label{eq:A4norm}
    \tilde{A}^4 = 
    \begin{dcases}
        \frac{2a}{b}\left[\exp\left(\frac{b^2}{4a}\right)K_{\frac{1}{4}}\left[\frac{b^2}{4a}\right]\right]^{-2} & \text{if}\,\, b>0
        \\
        -\frac{4a}{\pi^2b}\left[\exp\left(\frac{b^2}{4a}\right)\left(I_{-\frac{1}{4}}\left[\frac{b^2}{4a}\right]+I_{\frac{1}{4}}\left[\frac{b^2}{4a}\right]\right)\right]^{-2} & \text{if}\,\, b<0
        \\
        \sqrt{8a}\left[\Gamma\left(\frac{1}{4}\right)\right]^{-2} & \text{if}\,\, b = 0 
    \end{dcases} \,,
\end{align}
where $I_{\pm\frac{1}{4}}[z]$ and $K_{\frac{1}{4}}[z]$ are modified Bessel functions of the first and second kind, respectively. The normalization in the rescaled variables is $A^4 = \frac{1}{\sqrt{a}} \tilde{A}^4$.
\begin{align}
\label{eq:At4norm}
    A^4 = \frac{1}{\sqrt{a}} \tilde{A}^4 = 
    \begin{dcases}
        \frac{2}{c}\left[\exp\left(\frac{c^2}{4}\right)K_{\frac{1}{4}}\left[\frac{c^2}{4}\right]\right]^{-2}\,, & \text{if}\,\, c>0
        \\
        -\frac{4}{\pi^2c}\left[\exp\left(\frac{c^2}{4}\right)\left(I_{-\frac{1}{4}}\left[\frac{c^2}{4}\right]+I_{\frac{1}{4}}\left[\frac{c^2}{4}\right]\right)\right]^{-2}\,, & \text{if}\,\, c<0
        \\
        \sqrt{8}\left[\Gamma\left(\frac{1}{4}\right)\right]^{-2}\,, & \text{if}\,\, c = 0 \,,
    \end{dcases}
\end{align} 

The function $\tilde{f}$ can be defined piece-wise in terms of $\tilde{u}(\tilde{x}_1,\tilde{x}_1^{\prime}) \equiv 4b+3a(\tilde{x}_1^2+\tilde{x}_1^{\prime2})\,$
\begin{multline}
    \tilde{f}(\tilde{x}_1, \tilde{x}_1^\prime, a,b) =
\\
\begin{dcases}
\label{eq:fx1ab}
    \sqrt{\frac{\tilde{u}}{a}}
    \exp\left[\frac{\tilde{u}^2}{32a}\right]K_{\frac{1}{4}}\left[\frac{\tilde{u}^2}{32a}\right]  & \text{for} \,\, \tilde{x}_1^2+\tilde{x}_1^{\prime2} > -\frac{4b}{3a}
    \\
   \pi\sqrt{\frac{-\tilde{u}}{2a}}
    \exp\left[\frac{\tilde{u}^2}{32a}\right]\left(I_{-\frac{1}{4}}\left[\frac{\tilde{u}^2}{32a}\right] + I_{\frac{1}{4}}\left[\frac{\tilde{u}^2}{32a}\right] \right) & \text{for} \,\, \tilde{x}_1^2+\tilde{x}_1^{\prime2} < -\frac{4b}{3a}
    \\
     \sqrt{\frac{1}{2\sqrt{a}}}\,\Gamma\left(\frac{1}{4}\right) & \text{for} \,\, \tilde{x}_1^2+\tilde{x}_1^{\prime 2} = -\frac{4b}{3a}\,.
\end{dcases}
\end{multline}
 At the boundary, $\tilde{x}_1^2+\tilde{x}_1^{\prime2} = -\frac{4b}{3a}$, the value of expression $\tilde{f}(\tilde{x}_1, \tilde{x}_1^\prime)$ is dependent only on the value of $a$, regardless of the value of $b$, though the function is always piece-wise continuous. If $b>0$, the expression is simpler since it is always true that $\tilde{x}_1^2+\tilde{x}_1^{\prime2} > -\frac{4b}{3a}$. The relation to the rescaled variables is $f(x_1, x_1^\prime, c) = a^{\frac{1}{4}}\tilde{f}(\tilde{x}_1, \tilde{x}_1^\prime, a,b)$.

\subsubsection{Coupling non-identical Oscillators}
\label{app:nonid}

When $c_1$ and $c_2$ are sufficiently different, approximate analytic expressions exist for the even raw moments beyond the variance: 
\begin{equation}
\label{eq:rawmomx1}
    \mu_{2n}(x_1) = \expval{x_1^{2n}} \approx |\mu_{2n}(y_2) - \mu_{2n}(y_1)|\sin^2(\theta) + \min\{\mu_{2n}(y_1), \mu_{2n}(y_2)\}\,,
\end{equation}
\begin{equation}
\label{eq:rawmomx2}
    \mu_{2n}(x_2) = \expval{x_2^{2n}} \approx |\mu_{2n}(y_2) - \mu_{2n}(y_1)|\cos^2(\theta) + \min\{\mu_{2n}(y_1), \mu_{2n}(y_2)\}\,.
\end{equation}
This approximation successfully reproduced the expected shape for $\mu_{4}(x_1)$ when $c_1 = -1$, $c_2 = -5$ but performs quite poorly when $c_1 = -5.1$, $c_2 = -5$.
Then, for excess moments, when $c_1$ and $c_2$ are sufficiently different,
\begin{equation}
    \label{exmomx1}
    \nu_{2n}(x_1) \approx \frac{|\mu_{2n}(y_2) - \mu_{2n}(y_1)|\sin^2(\theta) + \min\{\mu_{2n}(y_1), \mu_{2n}(y_2)\}}{\left[|\text{Var}(y_2) - \text{Var}(y_1)|\sin^2(\theta) + \min\{\text{Var}(y_1), \text{Var}(y_2)\}\right]^n} - (2n-1)!!
\end{equation}
\begin{equation}
    \label{exmomx1}
    \nu_{2n}(x_2) \approx \frac{|\mu_{2n}(y_2) - \mu_{2n}(y_1)|\cos^2(\theta) + \min\{\mu_{2n}(y_1), \mu_{2n}(y_2)\}}{\left[|\text{Var}(y_2) - \text{Var}(y_1)|\cos^2(\theta) + \min\{\text{Var}(y_1), \text{Var}(y_2)\}\right]^n} - (2n-1)!!
\end{equation}
These approximations do well, for example, for $\nu_{4}(x_1)$ and $\nu_{6}(x_1)$ when $c_1 = -1$, $c_2 = -5$. But, they are less accurate for higher moments and for all moments when $c_1\approx c_2$, where they incorrectly predict $ \nu_{2n}(x_1) = \nu_{2n}(y_1)$.




\bibliographystyle{JHEP}
\bibliography{biblio.bib}






\end{document}